\global\def\draftcontrol{0}

   \def\versionno{ g2orbi -- atmp -- 1.7.02   }

\catcode`\@=11

\expandafter\ifx\csname draftcontrol\endcsname\relax\global\def\draftcontrol{0}
\fi

{\count255=\time\divide\count255 by 60
\xdef\hourmin{\number\count255}
\multiply\count255 by-60\advance\count255 by\time
\xdef\hourmin{\hourmin:\ifnum\count255<10 0\fi\the\count255}}
\def\draftdate{\number\month/\number\day/\number\year\ \ \ \hourmin }

\newcommand\makepapertitle{\par
  \begingroup
    \renewcommand\thefootnote{\@fnsymbol\c@footnote}%
    \def\@makefnmark{\rlap{\@textsuperscript{\normalfont\@thefnmark}}}%
    \long\def\@makefntext##1{\parindent 1em\noindent
            \hb@xt@1.8em{%
                \hss\@textsuperscript{\normalfont\@thefnmark}}##1}%
     \newpage
     \global\@topnum\z@   
     \@makepapertitle
     \thispagestyle{empty}\@thanks
  \endgroup
  \setcounter{footnote}{0}%
  \global\let\thanks\relax
  \global\let\makepapertitle\relax
  \global\let\@makepapertitle\relax
  \global\let\@thanks\@empty
  \global\let\@author\@empty
  \global\let\@date\@empty
  \global\let\@title\@empty
  \global\let\title\relax
  \global\let\author\relax
  \global\let\date\relax
  \global\let\and\relax
  \def\version{\let\version\@version\@gobble}
}
\def\@makepapertitle{%
  \newpage
   \ifnum\draftcontrol=1 {}
   \version\versionno
   \vskip 3em%
   \else
   \hfill\hbox to 3cm {\parbox{4cm}{\@pubnum}\hss}%
   \vskip 3em%
   \fi
   \begin{center}%
   \let \footnote \thanks
     {\LARGE \nointerlineskip\moveleft .45cm \hbox{\@title}\nointerlineskip\par}%
     \vskip 1.5em%
     {\normalsize
       \lineskip .5em%
       \begin{tabular}[t]{c}%
         \@author
       \end{tabular}\par}%
     \vskip 1em%
     {\@bstract}%
     \end{center}%
     \vskip .5em
     \@date%
   \par
}

\gdef\@pubnum{}
\def\pubnum#1{%
  \gdef\@pubnum{#1}}

\gdef\@bstract{}
\def\Abstract#1{%
  \gdef\@bstract{%
   \parbox{\textwidth-0pc}{%
   \centerline{\bf Abstract}\penalty1000
   \noindent
   \renewcommand\baselinestretch{1.0}
   {#1}}}
}

\def\ps@paper{\let\@mkboth\@gobbletwo%
     \ifnum\draftcontrol=1
	\def\@oddfoot{\hbox to \textwidth{\tiny \versionno \hfil\tiny\draftdate}%
	\hskip -\textwidth \hbox to \textwidth{\hfil\rm\thepage\hfil}}%
     \else\def\@oddfoot{\hbox to \textwidth{\hfil\rm\thepage\hfil}}
     \fi
     \let\@evenfoot\@oddfoot
}

\newenvironment{acknowledgments}{%
\vskip 3.25ex
\noindent {\bf Acknowledgments}
}

\def\@version#1{\ifnum\draftcontrol=1
\typeout{}\typeout{#1}\typeout{}
\vskip3mm\centerline{\hbox{\fbox{\normalsize{\tt DRAFT -- #1 -- }
                   {\draftdate}}}}\vskip3mm
\fi}
\let\version\@version
\long\def\eqlabel#1{\ifnum\draftcontrol=1
                    \tag@false  
                    \tag*{(\theequation) \hbox to -0.2cm{\hspace{0cm}\small{#1}\hss}}
                    \refstepcounter{equation} 
                    \edef\@currentlabel{\theequation}
                    \ltx@label{#1}          
                    \else
                    \label{#1}
                    \fi
                    }
\let\st@bibitem\@bibitem
\let\st@lbibitem\@lbibitem
\ifnum\draftcontrol=1
  \def\@bibitem#1{%
    \st@bibitem{#1}\a@@label{#1}\ignorespaces}
  \def\@lbibitem[#1]#2{%
    \st@lbibitem[#1]{#2}\a@@label{#2}\ignorespaces}
  \def\a@@label#1{%
    \gdef\a@lab{\smash{\normalfont\small#1}}
    \ifvmode
      \if@inlabel
        \global\setbox\@labels\hbox{%
          \llap{\a@lab\let\a@lab\relax
                \kern\@totalleftmargin\kern\marginparsep}%
          \box\@labels}%
      \fi
    \fi}
\fi

\documentclass[11pt,twoside]{article}
\usepackage{atmp}
\usepackage{amsmath,amssymb,array,calc,rotating,epsfig,psfrag}
\usepackage[nosort]{cite}
\def\cala         {{\cal A}}

\def\calc         {{\cal C}}
\def\cald         {{\cal D}}

\def\calh         {{\cal H}}

\def\caln         {{\cal N}}
\def\calo         {{\cal O}}

\def\cals         {{\cal S}}
\def\calt         {{\cal T}}

\def\calw         {{\cal W}}

\def\complex      {{\mathbb C}}

\def\projective   {{\mathbb P}}

\def\reals        {{\mathbb R}}
\def\zet          {{\mathbb Z}}

\def\del          {\partial}

\def\ee           {{\rm e}}
\def\ii           {{\rm i}}
\def\chain        {{\circ}}

\def\ie{{\it i.e.}}

\def\revise#1       {\raisebox{-0em}{\rule{3pt}{1em}}%
                     \marginpar{\raisebox{.5em}{\vrule width3pt\
                     \vrule width0pt height 0pt depth0.5em
                     \hbox to 0cm{\hspace{0cm}{%
                     \parbox[t]{4em}{\raggedright\footnotesize{#1}}}\hss}}}}

\newcommand\nxt[1]  {\\\fnxt#1}

\def\ZZ{\zet}
\def\RR{\reals}
\def\PP{\projective}
\def\RP{\RR\PP}
\def\N{\caln}

\def\phir{\varphi}
\def\phii{\Im(\phi)}
\def\zbar{{\bar z}}
\def\zr{{\zeta}}
\def\ybar{{\bar y}}
\def\yr{{\upsilon}}
\def\xbar{{\bar x}}
\def\xr{{\xi}}
\def\Wr{{\calw}}
\def\Wbar{{\overline W}}
\def\L{{\rm L}}
\def\R{{\rm R}}
\def\Phibar{{\bar \Phi}}
\def\chibar{{\overline\chi}}
\def\omegat{{\tilde\omega}}
\def\de#1#2{{\rm d}^{#1}\!#2\,}
\def\De#1{{\cald}#1\,}
\def\twin{{{\tr}_{\rm tw}(-1)^F}}
\def\unin{{{\tr}_{\rm un}(-1)^F}}
\def\Wilson{\Sigma}
\def\onecycle{\gamma}
\def\prodprime{\mathop{{\prod}'}}
\def\sumprime{\mathop{{\sum}'}}

\def\J{{\rm J}}
\def\Jcoset{{\rm J}_{\rm coset}}

\def\mimo{{\calc}}
\def\mimoorb{\calc^{\omega}}
\def\calao{\cala^{\omega}}
\def\Nto{{\mathfrak A}^{\omega}}
\def\un{{}^{\rm un}}
\def\us{{}^{\rm us}}
\def\tw{{}^{\rm tw}}
\def\plus{{}_{{}^+}}
\def\minus{{}_{{}^-}}
\def\plusminus{{}_{{}^\pm}}
\def\gsq#1#2{%
    {\scriptstyle #1}\square\limits_{\scriptstyle #2}{\,}} 
\def\sqr#1#2{{\vcenter{\vbox{\hrule height.#2pt  
 \hbox{\vrule width.#2pt height#1pt \kern#1pt
 \vrule width.#2pt}\hrule height.#2pt}}}}
\def\square{%
  \mathop{\mathchoice{\sqr{12}{15}}{\sqr{9}{12}}{\sqr{6.3}{9}}{\sqr{4.5}{9}}}}

\newcommand{\bouket}[1]  {{|#1\rangle\!\rangle}}

\newcommand{\bashi}[1]   {{\|#1\rangle\!\rangle}}

\def\yboxit#1#2{\vbox{\hrule height #1 \hbox{\vrule width #1
\vbox{#2}\vrule width #1 }\hrule height #1 }}
\def\fillbox#1{\hbox to #1{\vbox to #1{\vfil}\hfil}}
\def\ybox{{\lower 1.3pt \yboxit{0.4pt}{\fillbox{8pt}}\hskip-0.2pt}}

\def\comments#1{}

\def\half{{\frac12}}

\def\tr{\mathop{\rm tr}}
\def\Re{{\rm Re\hskip0.1em}}
\def\Im{{\rm Im\hskip0.1em}}

\def\lcm{{\rm lcm}}

\def\CA{{\cal A}}

\def\CL{{\cal L}}

\def\CN{{\cal N}}

\def\CW{{\cal W}}

\def\sgn{{\rm sgn\ }}

\def\II{\relax{I\kern-.10em I}}

\def\IZ{\relax\ifmmode\mathchoice
{\hbox{\cmss Z\kern-.4em Z}}{\hbox{\cmss Z\kern-.4em Z}}
{\lower.9pt\hbox{\cmsss Z\kern-.4em Z}}
{\lower1.2pt\hbox{\cmsss Z\kern-.4em Z}}\else{\cmss Z\kern-.4em
Z}\fi}
\def\IB{\relax{\rm I\kern-.18em B}}
\def\IC{{\relax\hbox{$\inbar\kern-.3em{\rm C}$}}}
\def\ID{\relax{\rm I\kern-.18em D}}
\def\IE{\relax{\rm I\kern-.18em E}}
\def\IF{\relax{\rm I\kern-.18em F}}
\def\IG{\relax\hbox{$\inbar\kern-.3em{\rm G}$}}
\def\IGa{\relax\hbox{${\rm I}\kern-.18em\Gamma$}}
\def\IH{\relax{\rm I\kern-.18em H}}
\def\II{\relax{\rm I\kern-.18em I}}
\def\IK{\relax{\rm I\kern-.18em K}}
\def\IP{\relax{\rm I\kern-.18em P}}

\def\inbar{\,\vrule height1.5ex width.4pt depth0pt}
\def\mod{{\rm\; mod\;}}

\font\cmss=cmss10 \font\cmsss=cmss10 at 7pt
\def\IR{\relax{\rm I\kern-.18em R}}

\def\id{{\it id}}
\def\BR{\RR}

\def\lp10{l_P^{10}}
\def\lp11{l_P^{11}}

\catcode`\@=12
\copyrightnotice{2002}{6}{207}{278}
\begin{document}


\title{Discrete Torsion in Singular $G_2$-Manifolds and Real LG}
\url{hep-th/0203272}


\moveleft .1\linewidth
\hbox{\parbox[t]{.35\linewidth}{
\author{Radu Roiban}
\address{Department of Physics \\ University of California \\
Santa Barbara, CA 93106, USA}
\addressemail{\footnotesize roiban@vulcan.physics.ucsb.edu}} 
\parbox[t]{.4\linewidth}{
\author{Christian R\"omelsberger}
\address{Department of Physics \\ and \\
CIT-USC Center for Theoretical Physics \\
University of Southern Calfornia \\
Los Angeles, CA 90089, USA}
\addressemail{\footnotesize roemel@citusc.usc.edu}}
\parbox[t]{.35\linewidth}{
\author{Johannes Walcher}
\address{Institute for Theoretical Physics \\
University of California \\
Santa Barbara, CA 93106, USA}
\addressemail{\footnotesize walcher@itp.ucsb.edu}}}

\markboth{DISCRETE TORSION IN SINGULAR $G_2$-MANIFOLDS} 
{\it R. ROIBAN, C. R\"OMELSBERGER and J. WALCHER}

\begin{abstract}
We investigate strings at singularities of $G_2$-holonomy
manifolds which arise in $\ZZ_2$ orbifolds of Calabi-Yau spaces times a
circle. The singularities locally look like $\RR^4/\ZZ_2$ fibered over a SLAG,
and can globally be embedded in CICYs in weighted projective spaces.
The local model depends on the choice of a discrete torsion in the fibration,
and the global model on an anti-holomorphic involution of the Calabi-Yau
hypersurface. We determine how these choices are related to each other by
computing a Wilson surface detecting discrete torsion. We then follow
the same orbifolds to the non-geometric Landau-Ginzburg region
of moduli space. We argue that the symmetry-breaking twisted sectors are 
effectively captured by real Landau-Ginzburg potentials. In particular, we
find agreement in the low-energy spectra of strings computed from geometry 
and Gepner-model CFT. Along the way, we construct the full modular data of 
orbifolds of $\N=2$ minimal models by the mirror automorphism, and give a 
real-LG interpretation of their modular invariants. Some of the models 
provide examples of the mirror-symmetry phenomenon for $G_2$ holonomy.
\end{abstract}

\cutpage





\section{Introduction and Summary}

\nocite{shva,figueroa}
In this paper, we study strings on $G_2$-holonomy spaces with orbifold singularities.
The examples we analyze are representable as $\ZZ_2$ quotients of Calabi-Yau
threefolds times a circle, and in certain cases are singular limits of smooth
$G_2$-manifolds.

Such $G_2$-holonomy spaces with singularities play a fundamental role in
phenomenologically relevant compactifications of M-theory to four dimensions,
see \cite{pato,acharya5,acharya10,amv,csu,csu2,atwi,witten79,acwi} and references 
thereto. In these references, it is shown how ADE-singularities in codimension 
$4$ give rise to non-abelian gauge symmetries \cite{acharya10,amv}, and extra 
isolated singularities, to chiral fermions \cite{csu2,acwi}, in the low-energy 
effective theory in four dimensions. The resulting dynamics can sometimes be solved 
and this has led to a number of interesting insights concerning geometric 
realizations of phase transitions in field theory.
\nocite{papi,kkp,gkp,cglp4,cglp5,bggg,cglp6,brandhuber,cglp7,wittendeconstr,ios,
friedmann}
\nocite{egsu,suya,blbr,rowa,egsu2,blbr2}
In the present paper, however, we will put aside these perspectives of $G_2$ 
holonomy, and rather try to understand certain aspects of stringy geometry 
associated with exceptional holonomy, following \cite{shva,figueroa,blbr,rowa,
egsu2,blbr2}.

To pose the basic problem that is addressed in this paper, we consider, as an
example, the Calabi-Yau hypersurface
\begin{equation}
Y = \{[x_i];x_1^8 + x_2^8 + x_3^8 + x_4^8 + x_5^2 = 0\} 
\eqlabel{Y}
\end{equation}
in the complex weighted-projective space $\PP_{11114}^4$. We also have in mind an
anti-holomorphic involution of $Y$ such as
\begin{equation}
\omega: x_i \mapsto \xbar_i \,,
\eqlabel{involex}
\end{equation}
and are interested in the quotient $X=\frac{Y\times S^1}{\omega}$, where $\omega$ acts
as \eqref{involex} on $Y$ and as inversion on the circle. The holonomy of $X$ is
strictly larger than SU$(3)$, and the next available Lie group on Berger's list
is $G_2$. We will losely refer to such $X$ as a $G_2$-holonomy space.

Compactification of the type II string on $X$ will lead in three dimensions to a
theory with $\N=2$ supersymmetry. In the large-volume limit, we can determine
the massless spectrum of this theory from the classical geometry, or actually just
the topology, of $X$. Let us focus on the symmetry-breaking twisted sectors of the 
orbifold. To obtain massless twisted strings, we would need $\omega$ to have fixed 
points. But $\omega$ acts freely on $Y\times S^1$, simply because there are no real 
points on $Y$! So in this example, we do not expect any massless strings from the 
twisted sector.

As we now let $X$ shrink in size, stringy effects become important and classical 
geometry is less useful. In the small-volume limit, a much better description is
in terms of Landau-Ginzburg theory \cite{gvw,vawa,lvw,vafa4,witten41}. For $Y$ at 
hand, the relevant LG model is given by the superpotential
\begin{equation}
W = x_1^8 + x_2^8 + x_3^8 + x_4^8 \,,
\eqlabel{W}
\end{equation}
where the $x_i$'s are complex $\N=2$ LG fields, and a $\ZZ_8$ orbifold is implicit.
This potential is related in the obvious way to the polynomial in \eqref{Y} by
integrating out the massive field $x_5$. Let us again look at the $\omega$-twisted
sector of the orbifold. For massless strings, the twisted boundary conditions
set the imaginary part of the $x_i$'s to zero, and we obtain the restriction
of $W$ to real $x_i$'s. It is not hard to determine the groundstates for
this real LG potential. One finds in particular the Witten index in the twisted
sector to be $\twin=1$, in clear contradiction to the geometric result, which was
$0$. \footnote{The reader might worry that the total Witten index should always 
be zero on a seven-dimensional manifold, and also that the $\ZZ_8$ orbifold has 
not been taken into account yet. In fact, the full orbifold group is non-abelian
and this is crucial for determining the spectrum. We will be much more careful 
with these issues below, see in particular section \ref{full}.}

Of course, our argument relies on Landau-Ginzburg theory with $\N<2$ supersymmetries,
and it can not be taken for granted that the correspondence with geometry extends
to this situation. However, since we are at the Fermat point in LG moduli space,
we can also use the exactly solvable Gepner-model CFT \cite{gepner3,gepner5} based 
on tensor products of $\N=2$ minimal models. Indeed, it has been found in 
\cite{rowa,egsu2} that there can be massless modes in the twisted sectors of the
corresponding orbifolds precisely if all levels of the minimal models are even.
The model corresponding to \eqref{Y} is an example of this \cite{blbr}, with
levels $(6,6,6,6)$. Hence, also the Gepner model seems to contradict the geometrical
result.

In fact, since the theory in three dimensions has only 4 supercharges, one might
also imagine that there is a superpotential with a non-geometric branch opening
up at the Landau-Ginzburg point. This, however, is to be ruled out by the basic 
result of Shatashvili and Vafa \cite{shva} that the extended chiral algebra 
associated with $G_2$ holonomy suffices to protect marginal operators in 
conformal field theory. The role that in the $\N=2$ situation is played by the 
U$(1)$ current is here taken over by the tri-critical Ising model. It generates 
the extension beyond $\N=1$ worldsheet supersymmetry, and can be used to show, 
relying on results of \cite{dixon}, that any marginal operator is exactly marginal.

The apparent discrepancy between geometry and the Gepner model was first poin\-ted 
out in \cite{blbr}. Actually, there is a related puzzle, also noticed in \cite{blbr},
which arises if $Y$ is replaced with the quintic. Indeed, all levels in the
Gepner model are then odd, and there is no twisted massless mode. However,
at large volume, the fixed point locus of the involution is a non-trivial
$\RP^3$ and an ``adiabatic argument'' would imply a massless vector multiplet
in three dimensions.

It was proposed in \cite{blbr} that a solution of these puzzles might be related
to the fact that the Gepner models typically lie on the $B=\frac12$ line in
K\"ahler moduli space, while the geometry naturally has $B=0$. In the $\ZZ_2$
orbifold these two branches become disconnected, and the spectra need not agree.
However, a satisfactory dynamical explanation of the lifting of modes has not
been given so far. In particular, one needs to explain why the extra modes
appear sometimes on the $B=\frac12$ (as for $X$), and sometimes on the $B=0$ 
branch (as for the quintic).

We will show that the discrepancies actually disappear after a careful analysis
of the orbifold action, in particular on the B-field. Indeed, there are several
topologically distinct orbifolds, both in the geometry and in the LG/Gepner
model. The spectra in the twisted sector depend on the model, but agree after a
proper identification of the orbifolds at large and small volume. The B-field,
both through the bulk Calabi-Yau space and through the orbifold in the form of
discrete torsion, plays a crucial role in the analysis.

We now summarize the main results of the paper. The basic observation that
will solve the above puzzle is that the involution \eqref{involex} can be twisted
by the phase symmetries of the defining equation in \eqref{Y}, \ie, $x_i\mapsto
\ee^{2\pi\ii M_i/8}\xbar_i\,$, and that for a suitable choice of phases, the fixed
point set is determined by the {\it real} equation
\begin{equation}
\pm \xr_1^8\pm \xr_2^8 \pm \xr_3^8 \pm \xr_4^8 \pm \xr_5^2=0 \,.
\eqlabel{Yreal}
\end{equation}
The topology of the fixed point set does depend on the signs in \eqref{Yreal} and
certainly does not always exclude massless twisted strings. We will describe
these possibilities in more details, and for more general models, in section
\ref{geoinv}. On the conformal field theory side, the existence of massless
fields is really a result from the representation theory of the chiral algebra,
which is the chiral algebra of the Gepner model divided by $\omega$. More 
precisely, the Ramond ground states associated with the massless fields appear
at the level of the indivdual minimal models. But this does not imply that these
fields are actually contained in any modular invariant built on this chiral 
algebra. We will see in sections \ref{lgorbi} and \ref{full} that there are 
indeed modular invariants in which the Ramond ground states are absent.

The technical core of our paper can be found in sections \ref{lomo} through
\ref{lgorbi}. Basically, the local geometry of the singularity is the fibration
of an $A_1$ singularity over a supersymmetric three-cycle. Similarly to
geometric engineering \cite{kmp,kkv,kmv}, we first compactify the IIA string on 
the ALE space, which leads to an $\N=(1,1)$ gauge theory in six dimensions, at 
a generic point on the Coulomb branch \cite{aspinwall10}. We then compactify 
further down to three dimensions. In order to preserve supersymmetry, the theory 
has to be twisted by a non-trivial R-symmetry connection \cite{bvs}. However, 
it turns out that there is the possibility of an additional discrete twist by 
a real line bundle that couples to the quantum symmetry of the ALE space at the 
orbifold point in its moduli space. In particular, this twist can lift massless 
modes that would have been expected from topological twisting.

In section \ref{glomo}, we show how this discrete twist, which we identify with 
discrete torsion \cite{vafa}, arises in the global model. Following suggestions 
by Sharpe \cite{sharpe9,sharpe10,sharpe11}, we detect the discrete torsion by 
computing a Wilson surface in the covering space of the orbifold, \ie, an 
integral $\int_{\hat\Wilson} B$, where $\hat\Wilson$ is the covering of a torus 
worldsheet inside $Y\times S^1$. The non-trivial contribution to this Wilson 
surface comes from a boundary gluing term, which we show is non-zero precisely 
because $B=\frac 12$ on the Calabi-Yau space. We perform explicit calculations
for three examples, the quintic, $\PP_{11114}^4[8]$, and $\PP_{11222}^4[8]$---%
some of them in appendix \ref{sslag}---but our methods are generalizable to
other models.

We then leave geometry for a while and turn to a detailed study of orbifolds
of $\N=2$ minimal models by antiholomorphic involutions. Having obtained the
full modular data of the chiral theories in section \ref{mimo}, we illustrate 
in section \ref{lgorbi} the somewhat surprising connection between the twisted 
sectors of these orbifolds and real Landau-Ginzburg potentials. This connection
will be the basic tool to compute the string spectrum on the $G_2$-spaces at 
small volume. We emphasize, however, that the details of sections \ref{mimo} 
and \ref{lgorbi}, except possibly subsection \ref{rlg}, are not essential for 
an understanding of the geometrical parts of the paper.

We are then finally ready in section \ref{full} for the study of the 
(non-abelian) Landau-Ginzburg orbifolds that describe the small-volume 
regime of our $G_2$-holonomy spaces. We introduce an index that counts the 
total number of ground states, and discuss the notion of Poincar\'e duality in 
this context. We derive the massless spectra of twisted strings in this framework 
and show that they agree with the geometrical results.

We end with a speculation concerning mirror symmetry for $G_2$-holonomy manifolds.
In \cite{shva} it was argued that mirror symmetry should be viewed as the 
inaptitude of conformal field theory to completely decipher the geometry of 
the target space. For $G_2$ holonomy, strings can only detect the sum of Betti 
numbers $b_2+b_3$, but not $b_2$ or $b_3$ independently. This is similar to 
the mirror-symmetry phenomenon for Calabi-Yau threefolds, in which the Hodge 
numbers $h_{11}$ and $h_{21}$ can be determined from string theory only up to 
their exchange. More generally, if we take into account that discrete fluxes 
can lift modes from the naively expected massless spectrum, we are led to 
classify under mirror symmetry any collection of classical geometries with or 
without discrete fluxes that yield isomorphic conformal field theories when 
probed with strings. We will show that this phenomenon indeed appears in the 
situations studied in this paper. 

For an example, let us return to the manifold $X$, and to the relation between 
geometry and Landau-Ginzburg model. The continuation from large to small volume 
or vice-versa involves integrating-out or integrating-in the massive LG field 
$x_5$. In so doing, the phase of the quadratic piece in the potential is not 
determined, since it can simply be removed by redefinition of $x_5$. However, 
after quotienting by the involution $\omega$, a sign in front of $\xr_5^2$, with 
$\xr_5$ now real, can not be removed by a real change of variables. And indeed, 
the two real sections of $Y$, given by $\sum_i \xr_i^8 \pm \xr_5^2=0$, 
respectively, have distinct topologies. For $\pm=+$ the fixed point set of 
$\omega$ is empty, while for $\pm=-$, it consists of two copies of the real 
projective space $\RP^3$. So we have precisely the situation in which two 
distinct classical geometries lead at small volume to indistinguishable theories.
Of course, to match the spectra, one has to take into account the discrete fluxes 
that thread the large-volume cycles for one choice of signs. It is clear that 
this sort of mirror symmetry is a rather common phenomenon in our context. More
examples will become clear in section \ref{full}, including ones with massless
modes in the twisted sector. It will be interesting to see if they can be
extended to full-fledged mirror symmetries.

\section{Orbifolds of $G_2$ Holonomy}
\label{geoinv}

String compactification to $3$ dimensions with minimal supersymmetry
requires the compactification space to have $G_2$ holonomy, just as
$4$ dimensions require SU$(3)$ holonomy. It is a natural question to
ask how much of the usual Calabi-Yau story can be extended to
$G_2$ holonomy \cite{shva}. An important step in this program is the
construction of examples of manifolds admitting $G_2$-holonomy metrics
\cite{brsa,gpp}. While most of the recent progress on this issue is being
made in the non-compact situation \cite{cglp4,cglp5,bggg,cglp6,brandhuber,cglp7}, 
interesting physics is likely to emerge with compact $G_2$-manifolds, also 
from the M-theory perspective \cite{witten79,wittendeconstr}. In a sense 
the simplest compact $G_2$-manifolds can be obtained from orbifolding
Calabi-Yau threefolds \cite{joyce}, as discussed in the CFT framework
in \cite{figueroa}.

\subsection{$G_2$-manifolds from Calabi-Yau spaces}

A 7-manifold with $G_2$ holonomy has a covariantly constant, so-called
associative, 3-form which locally looks like \cite{mclean}
\begin{equation}
\phi=
dx^{567}+dx^5(dx^{12}-dx^{34})+dx^6(dx^{13}+dx^{24})+dx^7(dx^{14}-dx^{23})\,.
\eqlabel{phi}
\end{equation}
More precisely, written in this way, $\phi$ distinguishes a $G_2$ subgroup of
SO$(7)$ as its isotropy group, and is hence covariantly constant if the holonomy
is in $G_2$. We may embed ${\rm SO}(4)={\rm SU}(2)\times {\rm SU}(2)$ into
SO$(7)$ by acting on the first four coordinates by a normal rotation $(2,2)$ and
on the last three coordinates as if they were anti-self-dual forms $(1,3)$.
Then \eqref{phi} shows that SO$(4)$ is a subgroup of $G_2$.

Another maximal subgroup of $G_2$ is SU$(3)$. If a Calabi-Yau space $Y$
admits an antiholomorphic involution $\omega$ as an isometry, then
$X=\frac{Y\times S^1}{\ZZ_2}$ with the $\ZZ_2$ action being the combined
action of $\omega$ and $x\mapsto -x$ on the circle is a manifold of $G_2$
holonomy. The associative 3-form is then
\begin{equation}
\phi=J\wedge dx+\Re(\Omega)\,,
\eqlabel{assform}
\end{equation}
where $J$ is the K\"ahler form on $Y$ and $\Omega$ is the holomorphic 
3-form. The phase of $\Omega$ is fixed up to multiplication with $-1$ 
by the requirement that $\Omega\mapsto\bar\Omega$ under $\omega$. The 
sign ambiguity of $\Omega$ can be fixed by reversing the orientation 
of the $S^1$ and the fact that $\phi$ is only defined up to a nonvanishing 
real factor.

Singularities in $X$ arise if $\omega$ has fixed points. By construction,
the fixed point locus in $Y$ is a special Lagrangian submanifold, $M$. The 
singular set of $X$ then consists of two copies of $M$ because of the two 
fixed points on $S^1$. The local geometry of $Y$ around $M$ is described by 
the normal bundle $NM$ of $M\subset Y$. The complex structure on $Y$
identifies the normal bundle $NM$ with the tangent bundle $TM$ by 
multiplication with the imaginary unit. Thus, the local structure of such
a singular locus in $X$ is $X_L=\frac{TM\times\BR}{\ZZ_2}$ with the 
$\ZZ_2$ acting on the $\BR^4$ fiber of $TM\times\BR$ by reflection 
at the origin. This is a singular $A_1$ fibration over $M$.

The massless spectrum of type IIA string theory on a $G_2$-manifold $X$
consists of three kinds of multiplets in three dimensions. The gravity
multiplet contains the graviton and the RR-1-form $C_\mu$. Like in four
dimensions, the dilaton sits in a universal multiplet, which is a chiral
multiplet in three dimensions. The remainder of the vector- and 
chiral-multiplet spectrum depends strongly on the choice of the $X$.

If $X$ were nonsingular, we would obtain the usual spectrum of string theory 
after Kaluza-Klein reduction. This leads to $b_2(X)$ vector multiplets, due
to the B-field and the RR-3-form, and $b_3(X)$ chiral multiplets, due to
the metric deformations and the RR-3-form.

With the help of electric-magnetic duality, massless abelian vector 
multiplets and chiral multiplets can be converted into each other. From the 
point of view of 10-dimensional electric-magnetic duality, only the exchange
of all vector and all chiral multiplets seems natural, but from the 
three-dimensional point of view, we might also think about dualizing 
individual multiplets. This is mirror symmetry in three dimensions 
\cite{inse,dhoo,fst,guto}. Mirror symmetry for $G_2$-manifolds is similar to 
this \cite{shva,acharya3,agva3,gyz}. As defined in \cite{shva}, $G_2$-manifolds 
are mirror to one another if the sigma models on them give rise to identical 
conformal field theories. This implies that $b_2+b_3$ must be constant within 
a mirror family (which typically has more than two members). So switching the 
geometric interpretation of the same conformal field theory typically entails 
the exchange of a chiral with a vector multiplet. 

In our example of $X=\frac{Y\times S^1}{\ZZ_2}$, the Betti numbers of $X$ can be
determined from the Hodge numbers of $Y$ and the action of $\omega$. For example,
$H^2(Y)$ splits into positive and negative eigenspaces of $\omega$, $H_+^2(X)$ and
$H_-^2(X)$, which are invariant when combined with the zeroth and and first
cohomology of $S^1$, respectively. The split of $H^3(Y)=H^{3,0} \oplus H^{2,1}
\oplus H^{1,2}\oplus H^{0,3}$ leads to equal dimensions of positive and negative
eigenspaces. This gives the Betti numbers from the untwisted sector
\begin{equation}
\begin{split}
b^{(u)}_2(X)&=h_{1,1}^+ \,, \\
b^{(u)}_3(X)&=h_{1,1}^-+h_{2,1}+1\,.
\end{split}
\end{equation}

The massless untwisted 3-dimensional spectrum of type IIA theory on $X$ is 
then given by $b^{(u)}_2(X)$ vector multiplets together with $b^{(u)}_3(X)$ 
chiral multiplets. As for the twisted sector, the ``shrunk 2-cycle'' of the 
$A_1$ fibers can be combined with the $0$- and $1$-cycles of the fixed point 
locus $M$. As we will see in section \ref{lomototw}, one gets ${\hat b}_0(M)$ 
vector and ${\hat b}_1(M)$ chiral multiplets, where ${\hat b}_i(M)$ are 
certain twisted Betti numbers of $M$. The chiral multiplets in the twisted 
sector correspond to blowup modes for the singular locus, whereas the scalars 
in vector multiplets correspond to the B-field flux through the shrunk $S^2$ 
of the $A_1$ fiber.

\subsection{Involutions and GLSM}\label{geoinvglsm}

Many examples of Calabi-Yau manifolds can be realized as complete 
intersections in weighted-projective spaces, and it is natural to ask 
what the resulting $G_2$-holonomy geometries are, both at large and
at small volume. The natural framework for this is the gauged 
linear sigma model (GLSM) \cite{witten41}. At large volume, the GLSM
reduces at low energies to the non-linear sigma model, while at small 
volume, we obtain a Landau-Ginzburg orbifold.

Such a gauged linear sigma model has a number of U$(1)$ gauge groups
and chiral fields $x_i$ with charges $w_i^{(a)}$ together with a
gauge invariant superpotential $W(x_i)$. A possible antiholomorphic 
involution $\omega$ has to preserve the U$(1)$ gauge invariance and the 
superpotential. In order to preserve the flat metric and the origin on 
the space of the $x_i$, the involution $\omega$ has to act by a unitary 
transformation as
\begin{equation}
\omega: x_i\mapsto M_{ij}\bar x_j\,.
\end{equation}
Gauge invariance requires $M_{ij}$ to be block diagonal, where the
U$(1)$ charges in one block are all the same. Furthermore, $M_{ij}$ can be 
'rotated' by U$(1)$ gauge transformations.

Further restrictions on $M_{ij}$ follow from the requirement that the
superpotential $W$ be 'invariant' under $\omega$,
\begin{equation}\eqlabel{eqnwinv}
W(M_{ij}\bar x_j)=\overline{W(x_j)}\,.
\end{equation}
We will not try to classify those antiholomorphic involutions, but
rather want to understand the ones where $M_{ij}$ is diagonal. There 
are in general more complicated involutions, which for example permute 
some $x_i$ with the same U$(1)$ charges \cite{papi}.

For simplicity, let us restrict ourselves to the case of a single
U$(1)$ gauge group, five chiral superfields $x_i$ with charges 
$w_i$ and one chiral superfield $p$ of charge $w=-\sum w_i$. We take the
superpotential to be
\begin{equation}
W=p\sum_i x_i^{h_i}\,.
\end{equation}
The antiholomorphic involution is of the form
\begin{equation}\eqlabel{eqnglsminv}
(x_1,\ldots,x_5,p)\mapsto (\rho_1\bar x_1,\ldots,\rho_5\bar x_5,\rho\bar p)\,,
\end{equation}
where the condition \eqref{eqnwinv} puts certain constraints on the phases
$\rho_i$,
\begin{equation}
\rho_i^{h_i}=\rho^{-1}\,.
\end{equation}

We can use gauge transformations to fix $\rho=1$. Then the $\rho_i$ are
$h_i$-th roots of unity. The involution (\ref{eqnglsminv}) can then be 
viewed as the involution $x_i\mapsto\bar x_i$ together with the discrete 
global symmetry $x_i\mapsto\rho_ix_i$ of the theory. It is, however, not 
true that we can always remove the phases $\rho_i$ by a symmetry 
transformation. Indeed, the involution \eqref{eqnglsminv} acts on 
$\rho_i^\prime x_i$ as $\rho_i^\prime x_i\mapsto\rho_i(\rho_i^\prime)^{-1}
\bar x_i$ or equivalently as $x_i\mapsto\rho_i(\rho_i^\prime)^{-2}\bar x_i$.
This shows that two involutions are related by symmetry if and only if the 
$\rho_i$ differ by even powers of an $h_i$-th root of unity. For each even 
$h_i$, this leaves two essentially different choices for $\rho_i$, whereas if 
$h_i$ is odd all choices of $\rho_i$ are equivalent. Furthermore, two choices 
are equivalent if they differ by a residual gauge transformation, \ie, 
$\rho_i\mapsto \ee^{2\pi\ii/h_i} \rho_i$ for all $i$ simultaneously.

The gauged linear sigma model can now be used to relate the action of 
the antiholomorphic involution $\omega$ in the Calabi-Yau phase to the 
action in the Landau-Ginzburg phase. In the Calabi-Yau phase we have a 
hypersurface given by the equation $\sum x_i^{h_i}=0$ in a weighted-projective 
space $\PP_{(w_1,\ldots,w_5)}^4$, and in the Landau-Ginzburg phase we get a 
$\ZZ_h$ Landau Ginzburg orbifold (where $h=\lcm (h_i)$ and $w_i=h/h_i$).
The involutions act in both limits in the obvious way by 
$x_i\mapsto\rho_i\bar x_i$.

\section{Local Model}
\label{lomo}

The $G_2$-manifold $X=\frac{Y\times S^1}{\ZZ_2}$ has singularities where
the anti-holomorphic involution $\omega$ has fixed points. The local model
for such a singular locus is $X_L=\frac{TM\times\BR}{\ZZ_2}$, where $M$ is
a supersymmetric 3-cycle. We have seen that this is a singular $A_1$
fibration over $M$. String theory is non-singular because a B-field
threading the shrunk 2-cycle gives non-zero mass to the branes wrapped around 
it \cite{aspinwall10}. This situation is very much reminiscent of geometric
engineering, where the fibration of ADE singularities over Riemann
surfaces is used to design quantum field theories in 4 dimensions.
In this context, an analysis based on topological twisting \cite{bvs} 
gives the right answer for the spectrum in 4 dimensions \cite{kmp}.
We will see that the present situation is somewhat more complicated.

\subsection{The Topological Twisting}\label{lomototw}

The low-energy theory for type IIA strings on the $\BR^4/\ZZ_2$ orbifold
is an $\CN=(1,1)$ U$(1)$ gauge theory on the 6-dimensional fixed plane,
coupled to the massless 10-dimensional type IIA fields. The SO$(4)$ 
R-symmetry\footnote{Since we are dealing with spinors, we should actually
talk about ${\rm Spin}(4)$. In the following, we will only write Spin$(4)$ 
if there could be some possible confusion.} can be identified with the 
rotation group transverse to the fixed plane and is gauged for this reason. 
The SO$(4)$ gauge fields are 10-dimensional (non-normalizable) gravitons 
which are polarized with one index in the transverse space and one index 
in the orbifold directions. The field content of the 6-dimensional (twisted) 
subsector is summarized in the following table.

\begin{center}
\begin{tabular}{|c|c|c|}
\hline
Field     & SO$(5,1)_t$  & SO$(4)_r={\rm SU}(2)_c\times {\rm SU}(2)_a$ \\
\hline
$A_\mu$   & $6$        & $(1,1)$ \\
$\phi$    & $1$        & $(2,2)$ \\
$\psi$    & $4$        & $(2,1)$ \\
$\lambda$ & $4^\prime$ & $(1,2)$ \\
\hline
\end{tabular}
\end{center}

The gauge boson $A_\mu$ is a RR-field and $\phi$ is in the NSNS-sector. 
The 6-dimensional action for the bosons has to be consistent with those 
symmetries and with supersymmetry. The kinetic terms of the 6-dimensional 
action are
\begin{equation}\eqlabel{kinlag}
\CA_{K}=
\int d^6x \sqrt{-g_6}\left[(D_\mu\phi)^2 + (F_{\mu\nu})^2\right]\,,
\end{equation}
where the derivatives are defined as
\begin{equation}
\begin{array}{c}
F_{\mu\nu}=\partial_\mu A_\nu-\partial_\nu A_\mu \,, \\
D_\mu\phi^i=\partial_\mu\phi^i+\omega^i{}_j{}_\mu\phi^j\,,
\end{array}
\end{equation}
with $\omega^i{}_j{}_\mu$ being the SO$(4)$ gauge connection which is
induced from the connection on the $A_1$ fibration. It has a particular
value determined by the supersymmetry condition solved by the geometry of
the particular fibration $X_L=\frac{TM\times\BR}{\ZZ_2}$.

The familiar topological twisting \cite{bvs,kmp} now instructs us to view
this solution of the supersymmetry condition as a result of the embedding
of the SO$(3)$ structure group of the tangent bundle $TM$ diagonally into the 
SO$(4)/\ZZ_2$ structure group of $X_L$. This also means that the R-symmetry 
connection is given in terms of the Levi-Civita connection on the special 
Lagrangian fixed cycle $M$. Under this split, the transformation properties 
of the different 6-dimensional fields are as follows.

\begin{center}
\begin{tabular}{|c|c|c|c|}
\hline
Field     & SO$(2,1)_t$ & SO$(3)_i$ & SO$(4)_r={\rm SU}(2)_c\times {\rm SU}(2)_a$ \\
\hline
$A_\mu$   & $3$       & $1$     & $(1,1)$ \\
$A_i$     & $1$       & $3$     & $(1,1)$ \\
$\phi$    & $1$       & $1$     & $(2,2)$ \\
$\psi$    & $2$       & $2$     & $(2,1)$ \\
$\lambda$ & $2$       & $2$     & $(1,2)$ \\
\hline
\end{tabular}
\end{center}

The diagonal topological twist retains four supercharges in three dimensions,
which are the singlets under the diagonal SO$(3)$. As for the field content, we 
have the singlets $(A_\mu,\phi^{(1)},\psi^{(1)},\lambda^{(1)})$ and the triplets
$(A_i,\phi^{(3)},\psi^{(3)},\lambda^{(3)})$. The singlets correspond to scalars 
on $M$, while the triplets would be one-forms in the usual assignement. Therefore, 
if we only take into account the topological twist, we predict that the 
3-dimensional $\CN=2$ theory has $b_0(M)$ vector multiplets and $b_1(M)$ chiral 
multiplets. In particular, this would give one vector multiplet for each connected 
component of $M$.

However, as remarked in \cite{blbr}, this result seems to be in contradiction 
to the results from the Gepner-model construction, where one obtains a rather
different spectrum of massless fields in the twisted sector. This indicates 
that the topological twist does not completely capture the topology of all 
fields involved in the compactification, and we should take a closer look
at possible subtleties.

On the one hand, we note that the $A_1$ fibration classically has an
SO$(4)/\ZZ_2$ structure group, but that the R-symmetry group is actually
SO$(4)$. So we have to address the question of existence and uniqueness
of this lift.

On the other hand, the theory with Lagrangian \eqref{kinlag} has, apart from
the gauge symmetries, that we discussed, a $\ZZ_2$ symmetry which leaves
invariant all the 10-dimensional fields and multiplies the twisted 6-dimensional
fields by $(-1)$. This is the $\ZZ_2$ quantum symmetry of the orbifold CFT.
It is unbroken only at the orbifold point, and it is broken in particular
at the point of enhanced SU$(2)$ gauge symmetry. In string theory, this
$\ZZ_2$ symmetry is gauged as well.

We now have to consider what the global field configuration is, including
all the continuous and discrete gauge symmetries\footnote{Note that the
U$(1)$ and the $\ZZ_2$ are not R-symmetries, whereas the SO$(4)$ is a
gauged R-symmetry.}.

We first look at the lift of the $A_1$ bundle to the SO$(4)$ R-symmetry 
bundle. The existence of the lift is determined by a second cohomology class 
$\tilde w_2$ analogous to the second Stiefel-Whitney class \cite{witten67,
witten60}. In the examples that we are studying, the local model is 
$X_L=\frac{TM\times\BR}{\ZZ_2}$ and we have the explicit lift to 
$TM\times\BR$. For this reason the class $\tilde w_2$ is vanishing. This 
lift is not necessarily unique, but the potential ambiguity in the lift 
to the R-symmetry bundle is fixed by the requirement of unbroken $\CN=2$ 
supersymmetry in 3 dimensions. Namely, the existence of unbroken 
supersymmetry requires the Spin$(3)_i$, the SU$(2)_c\,$, and the SU$(2)_a$ 
bundle all to be the same. This is also the reason why the ambiguity in the 
Spin$(3)$ bundle is irrelevant. Since all fields transform either 
trivially under the SO$(3)_i\times {\rm SU}(2)_c\times {\rm SU}(2)_a$ or as a
$(2,2,1)$, a $(2,1,2)$ or a $(1,2,2)$, a factor $(-1)$ in the Spin$(3)_i$ always 
gets squared to $1$.

For the $\ZZ_2$ quantum symmetry, gauging amounts to the choice of a real 
line bundle $L$ over $M$. If $L$ is nontrivial, the massless spectrum of 
twist fields on $M$ is no more described by the ordinary cohomology
$H^*(M,\BR)$. Because all 6-dimensional fields transform in the non-trivial
representation of the quantum $\ZZ_2$, the relevant cohomology is now the 
twisted cohomology $H^*(M,L)$, which is quite different as we will see below. 

Thus, from a physical point of view, the local model is given not only by
the special Lagrangian $M$, but also requires the choice of a real line 
bundle $L$. One might expect that this line bundle also plays a role from 
the mathematical point of view in the study of the resolvability of the 
orbifold singularities to smooth $G_2$-manifolds. It should then presumably 
also enter the proper definition of orbifold cohomology at the singularity.

\subsection{Gauging the quantum symmetry as a discrete torsion}
\label{lomodito}

Real line bundles on $M$ are classified by $H^1(M,\ZZ_2)$, which is the 
$\ZZ_2$ transition functions modulo equivalence\footnote{This is the first 
Stiefel-Whitney class $w_1$ of the real line bundle. This is similar to
complex line bundles, which are classified by the first Chern class. Actually,
the short-exact sequence $\ZZ\stackrel{\times 2}{\longrightarrow}\ZZ\stackrel
{\mod 2}{\longrightarrow}\ZZ_2$ induces a Bockstein homomorphism which maps 
$w_1$ to the first Chern class of a complex line bundle $\CL$, into which 
$L$ can be embedded. We will make use of this relation in section
\ref{retogeo}. Also note that in general $H^1(M,\ZZ_2)\ne H^1(M,\ZZ)\mod
2\,$!}. These transition functions also appear as the (discrete) holonomy
around a closed loop $\gamma\subset M$. For example, if there are no
non-trivial closed loops in $M$, then all real line bundles are trivial.

This gives a way to determine $L$ in string theory. We simply compute the
sign that a twisted string picks up when it propagates around a closed
loop $\gamma\subset M$. Such a twisted string propagating around $\gamma$ 
is a torus diagram embedded in $X_L$. Therefore, the holonomy of the real 
line bundle appears as a sign in front of a particular twisted partition 
function. In other words, we have a sign associated with a torus wrapping 
a non-trivial 2-cycle in $X_L\setminus M$, that is not determined by modular 
invariance. This is called discrete torsion \cite{vafa}.

In the local geometries that we are studying, this discrete torsion can
also be seen in a bit more conventional orbifold sense. In our examples
below, $M$ can be written as a $\ZZ_2$ quotient of some covering space $\tilde 
M$. Then the local geometry is $\frac{T\tilde M\times\BR}{\ZZ_2\times\ZZ_2}$, 
and the effect of turning on discrete torsion in this $\ZZ_2\times \ZZ_2$
orbifold corresponds to a non-trivial choice of $L$. In this case, the discrete
torsion can be understood in such geometrical terms as a choice of a bundle $L$
over $M$ because the $\ZZ_2$ acting on $\tilde M$ has no fixed points\footnote{%
Intriguingly, the double cover $\tilde M$ of $M$ seems to play a role from the
mathematical point of view \cite{joyce}.}.

This $\ZZ_2\times\ZZ_2$ orbifold of $\tilde M\times\BR$ can also be used to
calculate the twisted cohomology $H^*(M,L)$ from $H^*(\tilde M,\BR)$ by
projecting onto invariant forms. We will denote the dimensions of these
cohomologies (the twisted Betti numbers) by $\hat b_0$ and $\hat b_1$.

For example, consider $\tilde M=S^3=\{\zeta_1^2+\zeta_2^2+\zeta_3^2+
\zeta_4^2=0\}$ and the $\ZZ_2$ acting by the antipodal map
$(\zeta_1,\zeta_2,\zeta_3,\zeta_4)\mapsto(-\zeta_1,-\zeta_2,-\zeta_3,-\zeta_4)$,
and by $-1$ on the $\RR$-fiber, so that $L$ is non-trivial. Then, clearly,
$\hat b_0(M=\RP^3)=0$, and because $S^3$ had no 1-forms to begin with, also
$\hat b_1=0$.

As another example, consider $\tilde M=S^2\times S^1$, and the $\ZZ_2$
acting by the antipodal map on $S^2$ and inversion on the $S^1$. Then, if
$L$ is non-trivial, $\hat b_0$ is again $0$, but we now find $\hat b_1=1$.
If $L$ were trivial, we would have had $\hat b_0=1$ and $\hat b_1=0$.

This shows the dependence of the massless spectrum on $M$ and on the choice 
of the real line bundle $L$. We will see more examples in the next section.

\subsection{The M-theory picture}

One might ask the question what happens to the local model in the 
M-theory limit. In our understanding of the local model, the $\ZZ_2$ 
quantum symmetry is gauged and this gives rise to the discrete torsion.
In other words, the choice of discrete torsion corresponds to the choice 
of a real line bundle $L$ with the $\ZZ_2$ quantum symmetry as structure 
group.

More generally one might observe that in an ADE orbifold fibration,
one can gauge the discrete quantum symmetry group $Q$ and get a 
nontrivial principal $Q$ bundle over $M$. It is not hard to
see that the quantum symmetry group is the group of 1-dimensional
representations of the orbifold group. For ADE orbifolds, this
quantum symmetry group exactly agrees with the center ${\cal Z}$ of the 
enhanced ADE gauge group on the singularity.

One might then wonder whether there is any relationship between the
nontrivial $Q$ bundle in the string compactification and a discrete Wilson 
line in the M-theory lift \cite{wittendeconstr,ios,friedmann}. This, however, 
cannot be the case, since a discrete Wilson line which is in the center of 
the gauge group, does not have any effect on adjoint matter, whereas a 
Wilson line of the quantum symmetry group does.

The quantum symmetry is an exact symmetry of string theory
at the orbifold point, but it is spontaneously broken away from the 
orbifold point. In a lift to M-theory all nongeometric phases of 
a type IIA compactification are pushed away from the geometric
phases \cite{wittenph} and the point of enhanced SU$(2)$ gauge symmetry 
\cite{aspinwall10}, where as we saw, the $\ZZ_2$ quantum symmetry is 
broken due to terms involving the covariant derivative. This means that 
in the M-theory limit, the $\ZZ_2$ quantum symmetry is broken at a very 
high scale and actually disappears. It cannot be gauged anymore.

\section{Global models and Wilson Surfaces}
\label{glomo}

The local model for a singularity in our $G_2$-orbifolds is given as a
special Lagrangian three-cycle, $M$, which determines a singular $A_1$
fibration through topological twisting, plus the choice of a real line
bundle, $L$, over $M$. As we have discussed in the previous section, 
the spectrum of twisted strings at the singularity is determined by the
twisted cohomology groups $H^*(M,L)$.

The global models, on the other hand, are given as a Calabi-Yau threefold
$Y$ plus the choice of an anti-holomorphic involution, $\omega$. 

How does the global data determine the local model? It is not hard to find
the topology of the fixed point set, $M$, of $\omega$, and we will see
examples of this below. However, it is not clear {\it a priori} how to find
the line bundle $L$. For example, we do not expect discrete torsion to
be available in the global model. This is because, under certain assumptions,
discrete torsion in geometric orbifolds of the form $\hat X/\Gamma$, is
classified by $H^2(\Gamma,{\rm U}(1))$, and for $\Gamma=\ZZ_2$, this
cohomology group is simply trivial. The assumptions, explained in
\cite{sharpe11}, concern the fundamental group of $\hat X$, as well as the
torsion part of $H^2(\hat X,\ZZ)$. If the global model is, for example,
the quintic, then these assumptions are satisfied. They are, however,
clearly violated in the local model because $M=\RP^3$ has non-trivial
fundamental group. 

To reconcile this and to decide which line bundle to choose in the local model, 
we will calculate the relevant phase in the closed string torus amplitude for
a twisted string propagating around the non-trivial cycles of $M$, as
described in the previous section. It is not surprising that this phase will 
depend on the global B-field configuration on $Y$. Since B-fields, and in 
particular discrete ones, tend to be confusing, it is worthwile to clearly 
separate the following three kinds of B-fields that play a role in our 
discussion.

\begin{itemize}
\item In the local model, one can think of a B-field through the shrunk
$S^2$ in the $A_1$ fiber. This makes the CFT nonsingular, and the value
in the orbifold theory is $\half$. Excitations around this value are described 
by one of the four twisted scalar fields, denoted by $\phi^{(1)}$ in the 
previous section. For our considerations, it will not be important that
 $\phi^{(1)}$ can be interpreted as a B-field, and it is best to think of 
$\phi^{(1)}$ as simply one of the potentially massless fields after the 
topological twist.
\item The discrete torsion in the local model can be seen as a nontrivial
Wilson surface for the B-field \cite{vafa,sharpe11}. We will actually 
determine the discrete torsion by calculating this Wilson surface.
\item There are B-fields which are inherited from the moduli of the
Calabi-Yau space $Y$. Some of these B-fields are not invariant under the
$\ZZ_2$ orbifold action and are projected out as continuous moduli. But 
since the B-field is a cyclic variable \cite{witten20}, there are
two discrete invariant choices, $0$ and $\half$. We take this observation 
as our starting point.
\end{itemize}

\subsection{The torsion B-field in the global model}

We first draw a small cartoon of how the $\ZZ_2$ involution acts on the 
B-field. In a Calabi-Yau compactification of type IIA strings, the B-field 
behaves like an axion for the 4-dimensional abelian vector fields coming 
from the dimensional reduction of the RR-three form. If we further
compactify such a 4-dimensional gauge theory with axion $b$ on a circle,
we can consider dividing out a $\ZZ_2$ symmetry which reflects the
compactification circle and multiplies $b$ by $-1$.

\begin{figure}
\begin{center}
\psfrag{phia}{$b=\half$}
\psfrag{phib}{$b=-\half$}
\psfrag{sone}{$S^1$}
\psfrag{FP}{\parbox{2cm}{\footnotesize Fixed\\[-0.1cm] points}}
\epsfig{file=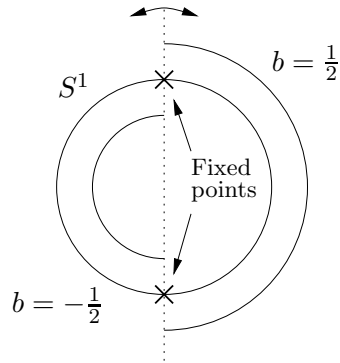,width=4cm}
\caption{The orbifold of the axion.}
\label{axion}
\end{center} 
\end{figure}

Because of the Witten effect \cite{witteneffect}, $b$ is a cyclic 
variable and can be fixed to two different values, $0$ and $\half$,
in the $\ZZ_2$ orbifold. If $b$ is fixed to $\half$ we can see from
a picture of the covering space (figure \ref{axion}) that the B-field
'jumps' at the fixed points of the $\ZZ_2$ orbifold. This suggests 
some interesting physics happening at the fixed points, which is,
however, hard to detect in the field theory.

In string theory, the B-field is a periodic variable because of gauge
symmetries. Recall that gauge transformations of the B-field are shifts
by integral closed two-forms. They can be encoded in 
a complex line bundle $\CL$, with gauge connection $\CA$. The B-field is 
then shifted by the field strength $F$ of $\CA$,
\begin{equation}
B\mapsto B+F \,,
\eqlabel{Bgt}
\end{equation}
which does not modify the field strength $dB$. More precisely, the
transformation \eqref{Bgt} is allowed at the level of closed strings which
(for topologically non-trivial configurations) only see the non-integral
part of the periods of $B$, while for open strings \eqref{Bgt} must be 
accompanied by a corresponding shift of the field strength on the branes.

As discussed at length in \cite{sharpe9,sharpe10,sharpe11}, these gauge
transformations give rise to interesting effects in the orbifold context.
Namely, if the orbifold acts non-trivially on the B-field, this can be a 
symmetry only when combined with a gauge transformation, which can affect 
topologically non-trivial configurations. This gauge transformation is 
precisely what happens to our axion in figure \ref{axion} at the fixed 
points of the orbifold.

More formally, line bundles are uniquely specified by their first Chern
class $c_1\in H^2(Y,\ZZ)$, which is the same as a field strength if
$H^2(Y,\ZZ)$ is torsion free. If the connection has moduli, these have to
be specified as well in order to define a unique gauge transformation
of the B-field. If $\pi_1(Y)$ vanishes, however, the connection is already
fully specified by the first Chern class. This is precisely the reason for
the two assumptions in \cite{sharpe11} that we have mentioned above.

In our examples, the action on the B-field and the accompanying gauge
transformations are determined geometrically. In the covering space 
$\hat X=Y\times S^1$, all B-fields are independent of the circle 
direction, which will be just a spectator in our analysis. In 
orientifold theories, for example, this $S^1$ might be omitted. 
Therefore, the orbifold action on the B-field is simply induced from
the action of $\omega$ on the second cohomology $H^2(Y,\ZZ)$, which
we know from the computation of the untwisted sector. For example, on
the quintic, we have $w\mapsto -w$ on the generator $w$ of the second
cohomology. So the B-field is projected out up to a discrete choice,
$B=0$ and $B=\frac{w}{2}$. The orbifold broke the non-torsion second
cohomology cycle of $Y$ to a $\ZZ_2$ torsion cycle on $X$.

\subsection{Calculation of the Wilson surface}

We are now in a position to calculate the Wilson surface for a twisted
string propagating around a 1-cycle, $\onecycle$, of a fixed special 
Lagrangian 3-cycle $M\subset Y$. We will relate this Wilson surface, which 
is discrete torsion in the local model, to the gauge transformation that 
in the global model is required to make the B-field invariant.

The Wilson surface, $\Wilson$, of our interest is a torus embedded in 
$X=\frac{Y\times S^1}{\ZZ_2}$. In fact, since we are interested in a
twisted string, $\Wilson$ descends from an annulus $\hat\Wilson$ in the
covering space $\hat X=Y\times S^1$. The two boundaries of $\hat\Sigma$ 
are glued together in $X$ by the orbifold action. 

The Wilson surface $\hat\Wilson$ receives contributions from the bulk of
$\Sigma$ as well as from the gluing of the boundaries \cite{sharpe11}. The 
bulk contribution vanishes because twisted strings are localized around the 
special Lagrangian, hence the annulus can be made arbitrarily narrow, and 
the B-field in the large-volume limit is very small. \footnote{One can 
actually embed $\Wilson$ in such a way in $\hat X$ that the integral explicitly 
vanishes.}

The boundary contributions to the Wilson surface are also described in
\cite{sharpe11}. Intuitively, we are inserting a gauge transformation
when gluing the boundary and this is simply a Wilson line of the line 
bundle describing the transformation. More properly, we may need 
several coordinate patches to define the B-field. The integrals of 
the B-fields in different coordinate patches have to be 'glued together' 
with Wilson lines of the line bundles describing the transition functions 
for the B-fields in neighboring coordinate patches. In our case, this
is the Wilson line of $\CL$ along $\onecycle$.

To relate this to the local model, we restrict $\CL$ to $M$. Because of the 
Lagrangian property, the restriction $\CL^\prime$ of the line bundle $\CL$ to 
$M$ has to be flat, \ie, the restriction of its first Chern class to $M$ has 
to be a torsion class in $H^2(M,\ZZ)$. Actually, one can extend the $\ZZ_2$ 
involution to $\CL$. The fixed point set of this involution is a real line 
bundle over $M$ which gives the holonomies of the flat line bundle $\CL^\prime$ 
around 1-cycles in $M$. This real line bundle is the line bundle $L$ considered 
in section \ref{lomodito}. These considerations not only show once more the 
relation of $L$ with discrete torsion, but also give an efficient way of
computing $L$ from global data. Moreover, it is also clear from this point 
of view that the existence of massless twisted fields is determined by the
twisted cohomology $H^\ast(M,L)$.

The same calculation for a Wilson surface also appears in a type IIA
string compactification on an orientifold of $Y$ which combines the
worldsheet orientation reversal with an antiholomorphic involution.
There this Wilson surface is an annulus amplitude between a D6-brane
wrapping the fixed cycle $M$ and its image. This changes now the open 
string spectrum on that D6-brane in a similar way.

\subsection{Real Toric geometry and some Examples}\label{retogeo}

We now return to the specific examples considered in section 
\ref{geoinvglsm}. That is, we assume that $Y$ in the large-volume limit 
is given as a Calabi-Yau hypersurface
\begin{equation}
\sum_{i=1}^5 x_i^{h_i}=0 \,,
\end{equation}
in the weighted-projective space $\PP_{w_1,\ldots,w_5}^4$. For simplicity,
we assume that this hypersurface is smooth, \ie, the singularities in the 
weigted-projective space are isolated.

Given the involution \eqref{eqnglsminv}, the equations for the fixed point 
set are
\begin{equation}\eqlabel{eqfixedptset}
[x_i]=[\rho_i \bar x_i]\,,
\end{equation}
together with the equation for the hypersurface. The restrictions on the
$\rho_i$, derived in section \ref{geoinvglsm} from gauge invariance in
the GLSM, here originate from invariance of the hypersurface equation
together with rescalings in $\PP_{w_1,\ldots,w_5}^4$.

The constraint \eqref{eqfixedptset} can be solved by setting
\begin{equation}
[x_i]=[\rho_i^\half\xi_i]\,,
\end{equation}
with $\xi_i\in\BR$. The sign ambiguity in $\rho_i^\half$ is removed
by the fact that $\xi_i$ can be positive or negative. This results
in the equation
\begin{equation}\eqlabel{eqrealhys}
\sum_{i=1}^5 \eta_i\xi_i^{h_i}=0\,,
\end{equation}
with $\eta_i=\bigl(\rho_i^\half\bigr)^{h_i}=\pm 1$, in the real
weighted-projective space $\RR\PP_{w_1,\ldots,w_5}^4$. Note that the 
solutions of equation \eqref{eqrealhys} only depend on the $\eta_i$ in 
front of even powers. We also note that the solutions of the real equation 
have to be modded out by a $\ZZ_2$ that is the real remnant of the U$(1)$ in 
the GLSM. More precisely, the $\ZZ_2$ acts by $\xi_i\mapsto -\xi_i$ for those 
$i$ with $w_i$ odd, and leaves all other $\xi_i$ invariant. 

A systematic way to solve for the special Lagrangian submanifold $M$ fixed 
by $\omega$ is to use real rescalings to solve eq.\ \eqref{eqrealhys} on 
a 4-sphere around the origin in $\RR^5$. This gives a double cover $\tilde 
M$ of $M$, which has to be modded out by the residual $\ZZ_2$. 

In order to determine the real line bundle $L$ over $M$, we describe 
the complex line bundle $\CL$ over $Y$ in terms of U$(1)$ charges. In 
the GLSM, the U$(1)$ charges of a section of $\CL$ are given by the first 
Chern class of $\CL$. This determines the action of the residual $\ZZ_2$
symmetry on the trivial real line bundle $\tilde M\times\BR$, yielding
$L=\frac{\tilde M\times\BR}{\ZZ_2}$. As in section \ref{lomodito}, the 
twisted cohomology can then be determined by looking at the $\ZZ_2$ action 
on the de Rham cohomology of $\tilde M$.

We now apply those techniques to a few examples. In section \ref{full}, we
will compare the geometrical results to results in the Landau-Ginzburg phase
of the GLSM.

\subsubsection{The Quintic}\label{secquintic}

The most popular Calabi-Yau threefold is the quintic in $\PP^4$. Its real
sections are determined by solving the real quintic equation on the $S^4$
in $\RR^5$, \ie, $\sum_{i=1}^5 \xi_i^5=0$. Over the reals, we can always
remove the $\eta_i$ in \eqref{eqrealhys}, so that they do not matter for
the topology of $\tilde M$. Also, we can uniquely define new variables 
$\zeta_i=\xi_i^5$. The (deformed) $S^4$ can then be written as $\sum 
\zeta_i^2=1$ and the quintic hypersurface is $\sum \zeta_i=0$. This shows 
that $\tilde M$ is a three sphere $S^3$.

The U$(1)$ charges of the homogeneous coordinates of $\PP^4$ and of
the section, $y$, of a complex line bundle are given by the table
\begin{equation}
\begin{array}{cccccc}
x_1 & x_2 & x_3 & x_4 & x_5 & y \\
 1  &  1  &  1  &  1  &  1  &  c_1(\CL) 
\end{array}
\end{equation}
Therefore, the residual $\ZZ_2$ symmetry acts by inverting all $\xi_i\,$,
and the special Lagrangian $M$ is $S^3/\ZZ_2=\RP^3$. The real line 
bundle $L$ is trivial if the first Chern class of $\CL$ is even, and 
has a nontrivial $\ZZ_2$ Wilson line if the first Chern class of $\CL$ 
is odd.\footnote{These special Lagrangians and the discrete Wilson line
have appeared, in a somewhat different context, in \cite{bbs,bdlr}.}

In particular, when comparing with results from Landau-Ginzburg and
Gepner models, we are in a situation with $c_1(\CL)$ odd. This is
because the real section of the K\"ahler moduli space passing
through the Landau-Ginzburg orbifold point originates with $B=\half$
in the large-volume region. (The section with $B=0$ in large volume 
goes through the conifold point and does not hit the Gepner point.) 
Therefore, $c_1(\CL)=w$, the K\"ahler class of the quintic.

So $L$ is non-trivial on the Gepner branch of the $G_2$ moduli space.
This shows immediately that $\hat b_0(M)=0$, and since $b_1(S^3)=0$,
we get $\hat b_1(M)=0$. According to section \ref{lomodito}, this leaves 
no massless twisted fields on the fixed cycle $M$.

\subsubsection{$\PP_{11114}^4[8]$}\label{secalleven}

\begin{table}[t]
\begin{center}
\begin{tabular}{|cc|c|c|c|cc|}
\hline
\#($\eta_i=-$) & \#($\eta_5=-$) & $\tilde M$ & 
$\ZZ_2$-action & $M$ & $\hat b_0$ & $\hat b_1$ \\
\hline
0 & 0 & $\emptyset$ & 
n/a & $\emptyset$ & 0 & 0 \\
0 & 1 & $S^3\times S^0$ & 
$(-{-}{-}-)(+)$ & $\RP^3\cup\RP^3$ & 0 & 0 \\
1 & 0 & $S^3\times S^0$ & 
$(+{-}{-}-)(-)$ & $S^3$ & 1 & 0 \\
1 & 1 & $S^2\times S^1$ & 
$(-{-}-)(+-)$ & $\frac{S^2\times S^1}{\ZZ_2}$ & 0 & 1 \\
2 & 0 & $S^2\times S^1$ & 
$(+{-}-)(--)$ & $\frac{S^2\times S^1}{\ZZ_2}$ & 0 & 0 \\
\hline
\end{tabular}
\caption{The fixed point sets on $\PP_{11114}^4[8]$, depending on the
combination of $\eta_i$. The third column shows how the residual $\ZZ_2$ 
acts on the coordinates of $S^p\times S^{3-p}$.}
\label{mm11114}
\end{center}
\end{table}

A second example is the degree $8$ hypersurface in $\PP_{11114}^4\,$,
\begin{equation}
x_1^8+x_2^8+x_3^8+x_4^8+x_5^2=0\,.
\end{equation}
Here all powers are even and we have to use a slightly different technique 
to determine $\tilde M$ depending on the choice of the $\eta_i$. We can 
leave all the terms with positive $\eta_i$ on the left side of the equation 
and bring all other terms to the right side. In this form both sides of 
the equation are positive definite. We can now impose the $S^4$ condition 
by setting both sides of the equation equal to some positive constant $R^2$. 
This clearly gives the product of a $p$-sphere and a $(3-p)$-sphere, 
$\tilde M=S^p\times S^{3-p}$.

The U$(1)$ charges of the homogeneous coordinates of $\PP_{11114}^4$ and the 
complex line bundle are given by the table
\begin{equation}
\begin{array}{cccccc}
x_1 & x_2 & x_3 & x_4 & x_5 & y \\
 1  &  1  &  1  &  1  &  4  & c_1(\CL)
\end{array}
\end{equation}
The B-field is again $\half$ and $c_1(\CL)=1$. Similarly to the quintic we 
can now determine $M$ and $L$. We have summarized the results in table 
\ref{mm11114}.

\subsubsection{$\PP_{11222}^4[8]$}\label{withblowup}

Our third example is the blowup of the degree $8$ hypersurface in
$\PP_{11222}^4$. The GLSM for the embedding space is given by the
charge table
\begin{equation}
\begin{array}{cccccc}
x_1 & x_2 & x_3 & x_4 & x_5 & x_6 \\
 0  &  0  &  1  &  1  &  1  &  1  \\
 1  &  1  &  0  &  0  &  0  & -2
\end{array}
\eqlabel{charget}
\end{equation}
and the hypersurface equation in homogeneous coordinates is \cite{kkp}
\begin{equation}
x_6^4 \bigl(x_1^8+x_2^8\bigr) + x_3^4+x_4^4 +x_5^4 = 0 \,.
\eqlabel{com11222}
\end{equation}

Because of the blowup, the solutions of the real versions of this equation
have to be subject to two independent rescalings and be modded out by a
$\ZZ_2\times\ZZ_2$ residual gauge symmetry. This is a little cumbersome,
and we have relegated the details of the calculation to the appendix
\ref{sslag}. For most combinations of $\eta_i$, however, some simplifications
occur, and the fixed point sets can be determined elementarily. One
combination for which this is not possible is $(\eta_1,\eta_2,\eta_3,\eta_4,
\eta_5)=(+,-,-,+,+)$, see the appendix for details. We summarize the fixed 
point data in table \ref{mm11222}. Some of these cases have also been 
discussed in \cite{kkp}.

\begin{table}[t]
\begin{center}
\begin{tabular}{|cc|c|cc|}
\hline
\#$(\eta_1,\eta_2=-)$ & \#$(\eta_3,\eta_4,\eta_5=-)$ & Fixed Point Set &  
$\hat b_0$ & $\hat b_1$ \\
\hline
0 & 0 & $\emptyset$ & 
0 & 0 \\
0 & 1 & ${S^1\times S^2}$ &     
0 & 0 \\
0 & 2  & $S^1\times S^1\times S^1$ & 
0 & 0 \\
0 & 3 & ${S^1\times S^2}$ &
0 & 0 \\
1 & 0 & $S^3$ &
1 & 0 \\
1 & 1 & see appendix \ref{sslag} &
0 & 1 \\
\hline
\end{tabular}
\caption{Inequivalent fixed point sets in $\PP_{11222}^4[8]$.}
\label{mm11222}
\end{center}
\end{table}

As an example, consider the fourth row in table \ref{mm11222}. The real 
section of \eqref{com11222} is
\begin{equation}
\xi_6^4\bigl(\xi_1^8 + \xi_2^8\bigr) = \xi_3^4+\xi_4^4+\xi_5^4 \,.
\end{equation}
Because $\xi_6$ can never vanish, we can rescale it to $1$, thereby
absorbing also one of the $\ZZ_2$ residual gauge symmetries. What
remains is equivalent $S^1\times S^2/\ZZ_2$, where the $\ZZ_2$ acts
only on the $S^1$ as the antipodal map. This leads to $\hat b_0=
\hat b_1=0$.

\section{Minimal model orbifolds}
\label{mimo}

In the foregoing sections, we have described an efficient way of computing
the spectrum of strings at orbifold singularities of $G_2$-manifolds ${\rm 
CY}\times S^1/\ZZ_2$, in the large-volume limit of the CY moduli space. Our 
goal in section \ref{full} will be to follow the same orbifolds to the 
small-volume regime, in particular to the Landau-Ginzburg orbifold region. 
This is intended first of all as an independent benchmark for the large-volume
results. Secondly, the consistency and simplicity of the results will verify 
the expectation, advocated in \cite{shva}, that $G_2$ compactifications of 
strings are similar in many respects to Calabi-Yau compactifications. 

We will show in section \ref{full} that the twisted spectrum of our 
orbifolds is simply computable in the Landau-Ginzburg orbifold phase by using 
the real LG potential as a Morse function. As usual in the LG/Gepner-model 
context, the orbifold procedure simply ensures space-time supersymmetry 
\cite{gepner3,gepner5,vafa4}, while the essence of the idea is already visible 
at the level of individual $\N=2$ minimal models \cite{vawa,lvw}. We will 
proceed similarly and first illustrate the connection in the simplest cases 
of ADE minimal models, in the following section \ref{lgorbi}. As a preparation, 
we will need some results about the conformal field theory of these minimal 
model orbifolds, in particular their modular transformation matrices. This is
the subject of the present section.

These charge-conjugation orbifolds of $\N=2$ minimal models are the elementary
building blocks of $G_2$-holonomy Gepner models \cite{blbr,rowa,egsu2}.
Parts of their modular data appear in particular in \cite{egsu2}, based on
earlier results of \cite{zafa,qiu2,raya2}. The modular data has also entered
the construction of B-type boundary conditions \cite{resc,bdlr,fkllsw,mms}.
Here, we fill-in certain missing entries of the modular S-matrix, associated
with fixed points.

We stress that the orbifolds in question are different from the ones that are 
usually studied in the context of Landau-Ginzburg theory \cite{vafa4}. The 
latter arise from dividing out (subgroups of) the group of scaling symmetries 
of the Landau-Ginzburg potential. In the conformal field theory limit, the 
orbifolded theories differ from the original ones only by a (simple-current)
modification of the modular-invariant partition function, while the symmetry 
algebra still contains the $\N=2$ super-Virasoro algebra. For example, it is 
well-known \cite{grpl} that for A-type minimal models, the $\ZZ_{h}$ orbifold 
yields an isomorphic (``mirror'') model with inverted left-moving U$(1)$ charge
(\ie, it corresponds to the charge conjugation modular invariant), and that for 
$h$ even, the $\ZZ_2\subset\ZZ_{h}$ orbifold corresponds to forming the D-type 
modular invariant.

In contrast, the orbifolds of present interest are chiral. They arise from
dividing out the $\ZZ_2$ mirror automorphism of the $\N=2$ super-Virasoro
algebra,
\begin{equation}
\omega: \quad L_n \mapsto L_n\,, \qquad
G^{\pm}_r \mapsto G^{\mp}_r\,, \qquad
J_n \mapsto -J_n \,.
\eqlabel{N2auto}
\end{equation}
In particular, this orbifold breaks $\N=2$ supersymmetry. Let us denote by $\Nto$ 
the subalgebra of the $\N=2$ superconformal algebra that is left pointwise fixed 
by \eqref{N2auto}. Our task in this section will be to investigate the 
representation theory of $\Nto$.

In fact, since our final goal is to write down modular-invariant partition functions
for minimal-model orbifolds and orbifolded Gepner models, all we need is the
modular data. This means establishing a list of primary fields, 
and finding their conformal weights and a matrix $S$ that describes the
modular-transformation properties of their characters. There are well-known
techniques to accomplish this task. We will mainly follow notations and
conventions of \nocite{dvvv}\cite{bifs}, and have summarized the relevant sections 
of this reference in appendix \ref{chzto}. Let us, however, mention that we are 
not rigorously studying the canonical representation theory of $\Nto$, which is a 
certain $\calw$-algebra \cite{blbr}. Such approaches in the context of $G_2$ 
holonomy have recently been taken in \cite{geno,noyvert}.

\subsection{The full modular data of $(\N=2)/\ZZ_2$}

Recall that the rational $\N=2$ superconformal algebras at central charge
$c=3k/h$, with $k\in\ZZ$ and $h=k+2$, have a realization as the chiral algebras 
of coset conformal field theories,
\begin{equation}
\mimo_k = \frac{{\rm SU}(2)_k\times {\rm U}(1)_4}{{\rm U}(1)_{2h}} \,.
\eqlabel{mimocoset}
\end{equation}
Accordingly, the (bosonic) primary fields are labelled by three integers
$l$, $m$, and $s$ with $0\le l\le k$, $m\in\ZZ_{2h}$, $s\in\ZZ_4$, subject
to the selection rule that $l+m+s$ be even and to the field identification
$(l,m,s)\equiv (k-l,m+h,s+2)$. There is a total of $2(k+1)(k+2)$ bosonic
primary fields, which can be organized in $(k+1)(k+2)$ primaries of the
$\N=2$ algebra. We wish to compute the orbifold of $\mimo_k$ by the chiral 
$\ZZ_2$ automorphism $\omega$, eq.\ \eqref{N2auto}. We will denote this CFT 
by $\mimoorb_k$. 

To begin with, it is easy to see that the action on primary fields induced
by $\omega$ is
\begin{equation}
\omega^*: \quad \phi_{(l,m,s)} \mapsto \phi_{(l,-m,-s)}\,.
\eqlabel{primauto}
\end{equation}
The symmetric fields, \ie, those fixed under $\omega^*$, are therefore 
precisely those with labels of the form $(l,0,0)$, $(l,h,0)$, $(l,0,2)$, 
$(l,h,2)$, and, for $k$ even, $(k/2,\pm h/2, \pm 1)$. It is important to keep 
in mind that the latter are symmetric only because of field identification 
in the coset construction. After field identification, there are then $k+1$ 
symmetric fields for $k$ odd, and $k+4$ for $k$ even. Each of these give rise 
to two primary fields of the orbifold. All others are pairwise identified by 
$\omega^*$ to give rise to primary fields with respect to $\Nto$.

\subsubsection{The strategy}

To proceed (see appendix \ref{chzto}), one has to compute the $k+1$ or $k+4$ 
twining characters, their modular S-transformations, and to decompose the results 
into the same number of characters $\chi^{(1)}$, which then yield the twisted 
characters. Explicit formulae for these characters can be found in refs.\
\cite{egsu2} and \cite{blbr}. However, similarly to many other situations of
this type, some of the twining characters actually vanish, and it is not possible
to compute the full modular data in this way. Furthermore, it is {\it a priori}
not clear what set of labels to use for the twisted sectors. But there is way
to circumvent these problems---at the price of others.

Given the coset representation \eqref{mimocoset}, it is quite natural to think of
the orbifold of our interest as a ``coset of orbifolds''. Namely, the
SU$(2)_k\times$U$(1)_4$ current algebra possesses a $\ZZ_2$ charge conjugation
automorphism which when restricted to the diagonal U$(1)_{2h}$ of the denominator
of \eqref{mimocoset} also becomes charge conjugation. The twining characters
of these algebras and automorphisms are well-known from the results of
\nocite{dvvv,bifs}\cite{dvvv,bifs}. We have summarized this data in appendix 
\ref{sutuoo}.

One can then proceed as in the usual coset construction and decompose the twining
and twisted characters of the numerator into those of the denominator. The
resulting branching functions, upon field identification and fixed point splitting,
then yield the desired character functions of $\mimoorb_k$. More formally, the
modular tranformation properties of the branching functions are obtained by
performing the appropriate simple-current projection on the tensor product of the
modular data of SU$(2)$ and U$(1)$ orbifolds. The only information that this method
does not give is the fixed-point-resolution prescription.

Quite generally, fixed-point resolution in simple-current constructions
\cite{scya6} requires the knowledge of a particular so-called fixed-point
S-matrix that describes the modular transformation properties of one-point
conformal blocks on the torus with simple-current insertions. These matrices are
known explicitly only for WZW models \cite{fss6}, and unfortunately not for their
(chiral) orbifolds. We therefore have to add yet another clue towards constructing
the charge conjugation orbifold of \eqref{mimocoset}.

It is by now well appreciated that modular data enters also in the description of
conformally invariant boundary conditions. In particular, for boundary conditions
that preserve the whole chiral algebra, the modular S-matrix is the change of basis 
that connects Ishibashi and boundary states (and hence closed and open string channel) 
\cite{cardy9}. Furthermore, it is known \cite{fusc11} that the modular data of the 
orbifold by a chiral automorphism $\omega$ yields the boundary conditions that
break the chiral symmetry algebra $\cala$ with definite automorphism type given by
$\omega$.

For an $\N=2$ minimal model, as for any $\N=2$ SCFT, boundary conditions
preserving the $\N=2$ algebra are usually called A-type, while those that
realize the $\N=2$ algebra by the mirror automorphism are said to be of
B-type. Accordingly, the modular data of $\Nto$ describes the change of
basis between Ishibashi and boundary states for B-type boundary conditions in
an (untwisted!) $\N=2$ minimal model. It would thus seem that constructing
these boundary conditions requires the modular data of $\Nto$ first. However,
these B-type boundary conditions can also be constructed by a different route!

Namely, B-type boundary conditions are mapped to A-type boundary conditions under
mirror symmetry and, for minimal models, mirror symmetry is achieved by the
Greene-Plesser orbifold construction. The Greene-Plesser orbifold \cite{grpl} is 
of simple-current type and rather simple to construct. In particular, the fixed
point resolution problem is reduced to the situation studied in \cite{fhssw},
and only known fixed point resolution matrices are required. Up to the fixed
points, the Greene-Plesser construction is implicit already in \cite{resc,bdlr},
see also \cite{brsc}. The strategy for fixed point resolution was followed 
explicitly in \cite{fkllsw} for B-type boundary conditions in Gepner models, 
and in \cite{mms} in a variety of other situations. 

However, the connection to symmetry breaking boundary conditions does not give
the full modular data either. For instance, the only pieces of the S-matrix that
enter are those that connect untwisted sectors with twisted sectors, \ie, the
matrix $S^{(0)}$. While the transformation from untwisted to untwisted sectors
is known {\it a priori} from the original theory, the S-matrix for twisted
sectors can be reconstructed from $S^{(0)}$ given the T-matrix, see eqs.\
\eqref{sotwtw},\eqref{po}. But the boundary conditions do not contain any
information at all about conformal weights in the twisted sectors.

Luckily, the combination of the information gained from viewing $\mimoorb_k$ as an 
orbifold of cosets (which covers all sectors, but misses the fixed points) with 
the information obtained from boundary CFT (which covers the fixed points, but 
misses the T-matrix and also the S-matrix from twisted sector to twisted sector),
yields the full solution, as we now describe.

\subsubsection{Primary fields}

Let us start by listing the primary fields of $\mimoorb_k$. In the untwisted sector,
the labels are inherited from $\mimo_k$, as we have described above. We label
the twisted sectors of $\mimoorb_k$ by two integers, $\lambda=0,\ldots,k/2$ and
$\sigma=0,1$, and, if $k$ is even and $\lambda=k/2$, a fixed point resolution
label $\eta=\pm$. The appearance of $\eta$ is intimately linked to the
existence of the symmetric field $(k/2,h/2,\pm 1)$ in the untwisted sector.
In addition, there is the usual $\ZZ_2$ character $\psi=\pm$ to distinguish the 
two primary fields in the same twisted sector. The full labels for primary fields 
in the twisted sectors are thus of the form $\tw(\lambda,\sigma)\psi$ for
$\lambda<k/2$ and $\tw(k/2,\sigma,\eta)\psi$. The total number of twisted
sectors is equal to the number of untwisted symmetric sectors.

There are two ways to think about the labelling scheme in the twisted sector.
The first, which we shall prefer, stems from the construction of $\mimoorb_k$ 
as coset of orbifolds. At the beginning of this construction, we have 
the labels $(\lambda,\mu,\sigma)$, where $\lambda=0,\ldots,k$ labels a 
twisted sector of SU$(2)/\ZZ_2$ and $\mu,\sigma=0,1$ label a twisted sector 
of U$(1)_{2h}$ and U$(1)_{4}$ respectively (see apendix \ref{sutuoo} for the
conventions). In the coset construction, they are subject to the selection rule 
$\lambda+\mu+\sigma$ even, which renders $\mu$ redundant, and to the 
identification $(\lambda,\mu,\sigma)\equiv(k-\lambda,\mu+k,\sigma)$, which has 
a fixed point for $k$ even $\lambda=k/2$ and hence leads to the degeneracy label 
$\eta$.

The alternate way of understanding the labelling comes from the relation
to B-type boundary conditions in minimal models. Here, the labels are
inherited from labels for A-type boundary conditions, \ie, $(L,M,S)$,
with $L=0,\ldots,k$, $M\in\ZZ_{2h}$ and $S\in\ZZ_4$, $L+M+S$ even,
and $(L,M,S)\equiv(k-L,M+h,S+2)$, by taking orbits under the Greene-Plesser
group $\ZZ_h\times \ZZ_2$, $(L,M,S)\equiv (L,M+2,S)\equiv (L,M,S+2)$.
These orbits are then one-to-one to the twisted sectors described above.
Again, for $k$ even and $L=k/2$, a fixed point arises, which can be
resolved according to \cite{fhssw}.

The labelling scheme might seem confusing, and it is not totally obvious how
to take a good section through the various identifications. We show one
possibility in table \ref{tab:mimoorbprims}.

\begin{sidewaystable}
\begin{center}
\begin{tabular}{lllllcc}
\multicolumn{2}{c}{sector} & \multicolumn{2}{l}{labels and range} & 
conformal weight & 
\multicolumn{2}{c}{number of fields} \\
&&&& $\bmod \ZZ$ & $k$ odd & $k$ even \\
\hline

\multicolumn{2}{c}{untwisted NS} \\
\cline{1-2} \\[-0.3cm]
\parbox{3cm}{} & non-symmetric & $\un(l,m,0)$ & 
\parbox[t]{4cm}{$l=0,\ldots,k$\\ $m=1,\ldots, h-1$ \\ $l+m$ even} & 
$\Delta(l,m,0)$ &
$\frac{(k+1)^2}{2}$ & $\frac{k (k+2)}{2}$ \\[1.2cm]
               & symmetric & \parbox[t]{2cm}{$\us(l,0,0)\psi$ \\ 
	    $\us(l,h,0)\psi$}, $\psi=\pm$ &
\parbox[t]{4cm}{$l=0,\ldots,k$ \\ $l$ even/odd} &
$\Delta(l,m,0)$ &
$2(k+1)$ & $2(k+2)$ \\[0.2cm]

\multicolumn{2}{c}{untwisted R} \\
\cline{1-2} \\[-0.3cm]
&non-symmetric & $\un(l,m,1)$ & 
\parbox[t]{4cm}{$0\le l<k/2$ \\ $m=-h+1,\ldots,h$} &
$\Delta(l,m,s)$ &
$\frac{(k+1)(k+2)}{2}$ & $\frac{k(k+2)}{2}$ \\
&& $\un(k/2,m,1)$ & $m=-h/2+1,\ldots h/2+1$ &
$\Delta(l,m,s)$ &
$0$ & $k/2$ \\
& symmetric & $\us(k/2,h/2,\pm 1)\psi$ & $\psi=\pm$ &
$\Delta(l,m,s)$ &
$0$ & $4$ \\[0.2cm]

\multicolumn{2}{c}{twisted (NS\&R)} \\
\cline{1-2} \\[-0.3cm]
&& $\tw(\lambda,\sigma)\psi$ & 
\parbox[t]{4cm}{$0\le \lambda< k/2$ \\ $\sigma=0,1$, $\psi=\pm$} &
$\Delta(\lambda) + \frac14\bigl(1\!-\!\psi(-1)^{\lambda\!+
\!k(\lambda\!+\!\sigma)}\bigr)$&
$2(k+1)$ & $2k$ \\
&& $\tw(k/2,\sigma,\eta)\psi$ & $\sigma=0,1$, $\eta=\pm$, $\psi=\pm$&
$\Delta(\lambda) + \frac14\bigl(1\!-\!\psi(-1)^{k/2}\bigr)$ &
$0$ & $8$ \\
\end{tabular}
\caption{Labels for primary fields in $\mimoorb_k$}
\label{tab:mimoorbprims}
\end{center}
\end{sidewaystable}

\subsubsection{Modular T-matrix}

The conformal weights and modular T-matrix can be determined from the coset 
construction. In the untwisted sector, they are modulo integers equal to the 
ones of the ordinary minimal models, \ie,  
\begin{equation}
\Delta = \Delta(l,m,s) = \frac{l(l+2) - m^2}{4h} +\frac{s^2}8 \bmod\ZZ \,. 
\eqlabel{Deltaun}
\end{equation}
In the twisted sectors, we obtain similarly, modulo half integers, 
\begin{equation}
\Delta = \Delta(\lambda) = \frac{c}{24} + \frac{(k-2\lambda)^2}{16h}\bmod\ZZ/2 \,,
\eqlabel{Deltatw}
\end{equation}
where $c=3k/h$ is the central charge of the minimal model. The value of the 
conformal weights in the rationals can be read off from the explicit character 
formulae given in \cite{egsu2}. For instance, in the untwisted sector, we have 
to bring $(l,m,s)$ to the ``standard range'' before we can apply the above 
formula. Moreover, in the twisted sector, there is a conventional choice of how 
to split up the twisted character $\chi^{(1)}$ into two irreducible characters. 
We have included the conformal weights, modulo integers, in table 
\ref{tab:mimoorbprims}.

\subsubsection{Modular S-matrix}

We now turn to the explicit formulae for the modular S-matrix. Recall that in
the ordinary minimal model, this matrix is given by
\begin{equation}
S_{(l,m,s),(l',m',s')} = \frac1h \,
\sin\pi\frac{(l+1)(l'+1)}{h}\,\ee^{2\pi\ii(mm'/2h - ss'/4)}
\end{equation}
Applying the formulae \eqref{soun} from the appendix, this readily yields
\begin{equation}
\begin{split}
S_{\un(l,m,s),\un(l',m',s')} &= \frac{2}{h} \,
\sin\pi\frac{(l+1)(l'+1)}{h}\,\cos 2\pi\Bigl(\frac{mm'}{2h}-\frac{ss'}{4}\Bigr)\\[0.2cm]
S_{\un(l,m,s),\us(l',m',s')\psi} &= \frac{1}{h} \,
\sin\pi\frac{(l+1)(l'+1)}{h}\,\ee^{2\pi\ii(mm'/2h - ss'/4)}    \\[0.3cm]
S_{\un(l,m,s),\tw(\lambda,\sigma)\psi} &= 
S_{\un(l,m,s),\tw(k/2,\sigma,\eta)\psi} = 0 \\[0.3cm]
S_{\us(l,m,s)\psi,\us(l',m',s')\psi'} &= \frac{1}{2h} \,
\sin\pi\frac{(l+1)(l'+1)}{h}\, \ee^{2\pi\ii(mm'/2h-ss'/4)}  \,.
\end{split}
\eqlabel{sunus}
\end{equation}
We now need information about the matrix $S^{(0)}$. The parts of $S^{(0)}$ that do
not involve fixed points are obtained by combining the $S^{(0)}$ matrices for SU$(2)$
and U$(1)$. This yields the following entries of $S$.
\begin{equation}
\begin{split}
S_{\us(l,m,0)\psi,\tw(\lambda,\sigma)\psi'} &= \frac{\psi}{\sqrt{2h}} \,
\sin\pi\frac{(l+1)(\lambda+1)}{h}\, \ii^{m-l} \,
\begin{pmatrix} 1 & 1 \\ (-1)^\lambda & -(-1)^\lambda \end{pmatrix} \\
S_{\us(k/2,h/2,s)\psi,\tw(\lambda,\sigma)\psi'} &= 0 \,,
\end{split}
\eqlabel{sustw}
\end{equation}
where rows and columns of the $2\times 2$ matrix are indexed by $m=0,h$ and 
$\sigma=0,1$, respectively. The standard formulae \eqref{sotwtw}, \eqref{po}, 
then also give the S-matrix elements in the twisted sector, excluding fixed 
points,
\begin{equation}
S_{\tw(\lambda,\sigma)\psi,\tw(\lambda',\sigma')\psi'} =
\frac{\psi\,\psi'\,\ii^{-\lambda-\lambda'}\,\ee^{-2\pi\ii k/8}}{\sqrt{2h}} \, 
\sin\pi\frac{(\lambda+1)(\lambda'+1)}{h} \, \delta_{\sigma,1-\sigma'} \,
\tilde s_{\lambda+\sigma,\lambda'+\sigma'} \,.
\eqlabel{stwtw}
\end{equation}
Here, $\tilde s$ is the matrix
\begin{equation}
\begin{pmatrix} 1+\ii^h & (-1)^h - \ii^{-h} \\ (-1)^h - \ii^{-h} & 1+\ii^h
\end{pmatrix} \,,
\eqlabel{stilde}
\end{equation}
originating from the U$(1)_{2h}$ part of the coset (see eq.\ \eqref{u1os} in the
appendix).

Finding the remaining entries of the S-matrix involves fixed point resolution.
We here follow the approach of \cite{scya6}, guided by the requirements that the
S-matrix be unitarity, symmetric, modular, and that the fusions be integer. Of
course, a more systematic explanation of fixed point resolution in orbifolds,
analogous to \cite{fss6} for ordinary WZW models, would be desirable. This is,
however, beyond the present scope.

Let us first explain the nature of the fixed points. In the formal tensor product
of orbifolds SU$(2)_k\times$U$(1)_4\times{\rm U}(1)_{2h}^*$, the labels 
$(k/2,h/2,\pm 1)$ are of type untwisted {\it non-symmetric}. Under the formal 
extension of this tensor product by the simple current $\Jcoset = \us(k,h,2)\plus$ 
implementing the coset construction, $\un(k/2,h/2,\pm 1)$ is fixed and gives rise 
to the two fields $\us(k/2,h/2,\pm 1)\plusminus\,$, which are untwisted {\it 
symmetric} in $\mimoorb_k$. Thus, the fixed point degeneracy label is the $\psi$ 
label for these fields. In the twisted sector, the tensor product has the fields
$\tw(k/2,\sigma)\psi$, which are also fixed under $\Jcoset$. They are resolved
into the fields $\tw(k/2,\sigma,\eta)\psi$. We thus see that we require a
$6\times6$ fixed point resolution matrix $S^{\Jcoset}$, subject to the usual
constraints \cite{scya6}.

Pieces of $S^{\Jcoset}$ can be found from the connection to B-type boundary 
conditions in ordinary minimal models. Namely, the Cardy coefficient of the 
Ishibashi state $\bashi{(k/2,h/2,s)}_{\rm B}$ in the boundary state $\bouket{(k/2,
\sigma,\eta)}_{\rm B}$ is essentially equal to the matrix elements
$S^{(0)}_{(k/2,h/2,s),(k/2,\sigma,\eta)}$. On the other hand, we know by the 
usual fixed point resolution formula that
\begin{equation}
S_{\us(k/2,h/2,s)\psi,\tw(k/2,\sigma,\eta)\psi'} = \frac12 \bigl( 
\psi\eta \, S^{\Jcoset}_{(k/2,h/2,s),(k/2,\sigma)\psi'} \bigr) \,.
\end{equation}
Note that the S-matrix before resolution vanishes, because before extension,
the field $(k/2,h/2,s)$ is non-symmetric. Combining these facts with eq.\
\eqref{soustw}, and consulting \cite{fkllsw,fhssw,mms} for the B-type boundary
conditions, we then find
\begin{align}
S^{(0)}_{(k/2,h/2,s),(k/2,\sigma,\eta)} &= \frac\eta2 
\begin{pmatrix} 1 & 1 \\ -1 & 1 \end{pmatrix} \,, \\
\intertext{with rows and columns indexed by $s$ and $\sigma$, respectively. 
This finally yields}
S_{\us(l,m,0)\psi,\tw(k/2,\sigma,\eta)\psi'} &= \frac{\psi}{2\sqrt{2h}} \,
\sin\frac{\pi}{2}(l+1) \, \ii^{m-l} \,
\begin{pmatrix} 1 & 1 \\ (-1)^\lambda & -(-1)^\lambda \end{pmatrix} \notag\\
S_{\us(k/2,h/2,s)\psi,\tw(k/2,\sigma,\eta)\psi'} &= 
\frac{\psi\,\eta}{4}
\begin{pmatrix} 1 & 1 \\ -1 & 1 \end{pmatrix} \,.
\eqlabel{sustwf}
\end{align}

Now, the remaining elements of the S-matrix can be computed from the formulae
\eqref{sotwtw}, \eqref{po}, and we find
\begin{equation}
\begin{split}
S_{\tw(k/2,\sigma,\eta)\psi,\tw(\lambda',\sigma')\psi'} &=
\frac{\psi\,\psi'\,\ii^{-k/2-\lambda'} \, \ee^{-2\pi\ii k/8}}{2 \sqrt{2h}} \, 
\sin\frac{\pi}{2}(\lambda'+1) \, \delta_{\sigma,1-\sigma'} \,
\tilde s_{k/2+\sigma,\lambda'+\sigma'} \\
S_{\tw(k/2,\sigma,\eta)\psi,\tw(k/2,\sigma',\eta')\psi'} &=
\frac{\psi\,\psi'\, \ii^{-k} \, \ee^{-2\pi\ii k/8}}{4 \sqrt{2h}} \, 
\sin\frac{\pi}{2}\bigl(\frac{k}2\!+\!1\bigr) \, \delta_{\sigma,1-\sigma'} \,
\tilde s_{k/2+\sigma,k/2+\sigma'} \\
& \qquad\qquad\qquad\qquad\qquad\qquad +
\frac14 \,\psi\psi'\,\eta\eta' \, \delta_{\sigma,\sigma'} \,.
\end{split}
\eqlabel{stwtwf}
\end{equation}
Here, $\tilde s$ is given by \eqref{stilde}, and one may recognize
$\frac12\,\psi\psi'\,\delta_{\sigma,\sigma'}$ as the remaining entries of the
fixed point resolution matrix $S^{\Jcoset}$.

One can check that the S-matrix given by eqs.\ \eqref{sunus}, \eqref{sustw}, 
\eqref{stwtw}, \eqref{sustwf}, and \eqref{stwtwf} is unitary, satisfies
$(ST)^3 = S^2$, and yields integers upon insertion in the Verlinde formula.
Let us point out a few more aspects of the modular data that will be useful
for the following section.

\subsection{$\mimoorb_k$ as an $\N=1$ theory}
\label{aspects}

As we have mentioned above, the (bosonic) orbifold $\mimoorb_k$ has
$\N=1$ supersymmetry. The supercurrent is the (bosonic) primary
$v=\us(0,0,2)\plus\,$. In the untwisted sector, NS and R sectors are
distinguished by $s=0,2$ and $s=\pm 1$, respectively. In the twisted
sector, by looking at the monodromy of $v$, one can deduce that $\sigma=0$
corresponds to the R sector, and $\sigma=1$ to the NS sector.

As usual in the context of $\N=1$ theories, NS super-primaries come from
two bosonic primaries. For example, in the twisted sector, the fields
$\tw(\lambda,1)\plus$ and $v*\tw(\lambda,1)\plus=\tw(\lambda,1)\minus$ are
each others superpartners. In the Ramond sector, super-primaries usually
correspond to only one bosonic primary. For example, we have the fusion
rule $v*\un(l,m,1)=\un(l,m,1)$. But there are also cases in which a Ramond
super-primary does split into two bosonic primaries, for instance if there
is a ground state, with lowest conformal weight $\Delta= \frac c{24}\,$. 
From the formulae \eqref{Deltaun} and \eqref{Deltatw} for the conformal weights, 
one deduces that there are Ramond ground states only if $k$ is even, with 
$\lambda=k/2$. Arising from a fixed point, the $\eta$-label of this field 
is ambiguous (this is known as ``fixed point homogeneity'' \cite{fss6}).
A natural choice is to label the ground state by $\tw(k/2,0,+)\plus\,$.
Its worldsheet superpartner is $\tw(k/2,0,-)\plus\,$, with conformal
weight $\frac c{24}+1$. There are also examples (not for $\mimoorb_k\,$!) in 
which a Ramond field with $\Delta=\frac c{24}$ is its own superpartner. This
indicates that there are actually two ground states, with opposite
chirality.

Another question in $\N=1$ theories is the chirality of the Ramond ground states.
If we are interested in a non-chiral fermion number $(-1)^F$, we can answer
this question only at the level of the torus partition function including left-
and right-movers. We have the following convention. If in the bosonic partition
function, a R field with $\Delta=\frac c{24}\,$, such as $\tw(k/2,0,+)\plus\,$,
is paired with itself, the chirality of the corresponding ground state is
$(-1)^F=+1$. If it is paired with its superpartner $\tw(k/2,0,-)\plus\,$, the
chirality of the ground state is $(-1)^F=-1$.

\section{Real Landau-Ginzburg and minimal models}
\label{lgorbi}

It is well-known \cite{qiu,raya} that $\N=2$ minimal models are ADE classified
by the simply-laced finite-dimensional Lie algebras, $A_n$ for $n=1,2,\ldots$,
$D_n$ for $n=3,4,\ldots$, and $E_6$, $E_7$, and $E_8$. From the point of view of
conformal field theory, this is inherited from the famous Cappelli-Itzykson-Zuber 
ADE classification of modular invariants for SU$(2)$, see refs.\ \cite{ciz,ciz2}. 
From the point of view of effective Landau-Ginzburg theory, it is the classification 
of quasi-homogeneous holomorphic superpotentials with modality zero, and in 
particular is the basic link between the classification of conformal theories and 
singularity theory \cite{vawa}.

Through the Landau-Ginzburg description of $\N=2$ minimal models, the ADE
classification of modular invariants becomes equivalent to the ADE classification
of simple complex singularities. Since these singularities can also be written
as $\complex^2/\Gamma$, where $\Gamma$ is a finite subgroup of SU$(2)$, this
is also intimately connected to the ADE classification of finite subgroups of
SU$(2)$. Besides the classification, the correspondence manifests itself mainly
in certain combinatorial data associated with ADE. For example, the exponents
of the Lie algebras appear in the diagonal terms of the modular-invariant
partition functions, the local ring of the singularity is isomorphic to the
chiral ring of the superconformal model, the Coxeter element of the Weyl group
is a symmetry of the LG superpotential, {\it etc.}. A more recent example is the
realization \cite{lerche5,lls} that the ADE Dynkin diagram and the finiteness
of its root system is also contained in the conformal field theory, namely in
the classification of A-type boundary conditions and their intersection
properties.

In this section, we will add to this web of relations a link between orbifolds
of $\N=2$ superconformal models by antiholomorphic involutions and the
classification of {\it real} singularities, see \cite{agv}, chapter 17. 
Specifically, we will argue that the twisted sectors of the $(\N=2)/\ZZ_2$ minimal
models are governed by the real simple singular functions, just as the ordinary
$\N=2$ models are governed by the complex ones. To support this, we will
contruct modular invariants for the theories $\mimoorb_k$ considered in
the previous section, read off the supersymmetric index ${\rm tr}(-1)^F$ in the
twisted RR sector, and see that it agrees precisely with the Morse index of (the
deformation and stabilization of) the corresponding real singular function.

In the end, this link might not be totally surprising. In particular,
it turns out that the modular invariants for $(\N=2)/\ZZ_2$ can be obtained
by suitably twisting the orbifold action \eqref{N2auto} in the modular
invariants for $\N=2$ minimal models. This then parallels the fact \cite{agv}
that (at least for low modality) the real singularities are classified by
the possible real forms of the complex ones.

On the other hand, our results fill a much-needed gap in the LG-CFT connection,
by extending it to a less supersymmetric situation. It is known that there is 
a relation between $\N=1$ minimal models and $\N=1$ LG models (see, for 
instance \cite{kms}), and indeed the initial proposal of Zamolodchikov
\cite{zamolodchikov5} is concerned with $\N=0$. But the absence of
non-renormalization theorems for the superpotential for $\N<2$ makes these
relations much weaker than for $\N=2$. For instance, it is already hard
to see how the modular invariants for $\N=1$ minimal models \cite{cappelli}
are classified by $\N=1$ LG superpotentials \cite{vawa}. This correspondence
is much more explicit in our situation. It would be interesting to understand 
whether our results can be interpreted in the sense of some non-renormalization 
theorems for $(\N=2)/\ZZ_2$.

\subsection{$\N=2$ modular invariants}

Let us start by recalling the modular invariants for ordinary $\N=2$ 
minimal models. 

First of all, there is the diagonal modular invariant, also known as
A-type. It reads, for any $k\in\ZZ$,
\begin{equation}
Z^{A_{k+1}} = \sum_{(l,m,s)} \bigl| \chi_{(l,m,s)} \bigr|^2 \,,
\eqlabel{A}
\end{equation}
where the sum is over all allowed combinations $(l,m,s)$ modulo field
identification, \ie, $l=0,\ldots,k$, $m=0,\ldots,2h-1$, $s=0,1,2,3$
with $l+m+s$ even and $(l,m,s)\equiv(k-l,m+h,s+2)$. We will in general 
not specify the summation ranges as explicitly, since this is usually
quite cumbersome, but obvious from the context.

The D-type models exist for any even $k$. They can be understood as 
$\ZZ_2$ orbifolds\footnote{We here understand orbifolds in the string 
theory sense \cite{dhvw}. They can correspond, in the context of
rational conformal field theory, to chiral orbifolds, simple-current 
extensions, simple-current induced fusion-rule automorphisms, or a 
mixture of these.} of the A-type models, where the orbifoldization by 
$\J=(-1)^l$ projects onto integer spin. In other words, we have the 
twisted partition functions,
\begin{equation}
\begin{split}
\gsq{\J}{\id} &= \sum_{(l,m,s)} (-1)^l \bigl| \chi_{(l,m,s)} \bigr|^2 \\
\gsq{\id}{\J} &= \sum_{(l,m,s)} \chi_{(l,m,s)}\chibar_{(k-l,m,s)} \\
\gsq{\J}{\J}&= \sum_{(l,m,s)} (-1)^{k/2-l} \chi_{(l,m,s)} 
\chibar_{(k-l,m,s)} \,.
\end{split}
\end{equation}
The resulting partition function, denoted by $D_{k/2+2}$, is
\begin{align}
Z^{D_{k/2+2}} &=
\textstyle\frac12\displaystyle
\Bigl(\gsq{\id}{\id}+\gsq{\J}{\id}+\gsq{\id}{\J}+\gsq{\J}{\J}\Bigr)\notag\\
\text{$\textstyle\frac k2$ even:} \qquad\qquad &= \sum_{(l,m,s),\;l<k/2\;{\rm even}}
\bigl|\chi_{(l,m,s)} + \chi_{(k-l,m,s)}\bigr| ^2 
+ \sum_{(m,s)} 2 \,\bigl|\chi_{(k/2,m,s)}\bigr|^2 \eqlabel{Deven}\\
\text{$\textstyle\frac k2$ odd:} \qquad\qquad &= \sum_{(l,m,s),\;l\;{\rm even}}
\bigl|\chi_{(l,m,s)}\bigr|^2 +
\sum_{(l,m,s),\;l\;{\rm odd}} \chi_{(l,m,s)} \chibar_{(k-l,m,s)} \,,
\eqlabel{Dodd}
\end{align}
where we have made explicit that the $D_{\rm even}$ invariants are of
extension type, while the $D_{\rm odd}$ invariants are of pure automorphism
type. Both of them are simple-current modular invariants, constructed
from the simple current $\J=\phi_{(k,0,0)}\,$.

The exceptional modular invariants cannot be written as orbifolds. They
occur at level $k=10, 16$, and $28$ for $E_6$, $E_7$, and $E_8$,
respectively, and read
\begin{align}
Z^{E_6} &=
\sum_{(m,s)} \bigl|\chi_{(0,m,s)}+\chi_{(6,m,s)}\bigr|^2 +
 \bigl|\chi_{(4,m,s)}+\chi_{(10,m,s)}\bigr|^2 +
 \bigl|\chi_{(3,m,s)}+\chi_{(7,m,s)}\bigr|^2
\eqlabel{E6} \\[0.2cm]
Z^{E_7} &=
\sum_{(m,s)} \bigl|\chi_{(0,m,s)}+\chi_{(16,m,s)}\bigr|^2 +
\bigl|\chi_{(4,m,s)}+\chi_{(12,m,s)}\bigr|^2 +
\bigl|\chi_{(6,m,s)}+\chi_{(10,m,s)}\bigr|^2 
\notag\\[-0.2cm] &\qquad +
\bigl|\chi_{(8,m,s)}\bigr|^2 +
\bigl(\chi_{(2,m,s)}+\chi_{(14,m,s)}\bigr)\chibar_{(8,m,s)}
+\chi_{(8,m,s)}\bigl(\overline\chi_{(2,m,s)}+\chibar_{(14,m,s)}\bigr)
\eqlabel{E7} \\[0.2cm]
Z^{E_8} &=
\sum_{(m,s)} 
\bigl|\chi_{(0,m,s)}+\chi_{(10,m,s)}+\chi_{(18,m,s)}+\chi_{(28,m,s)}\bigr|^2
\notag\\[-0.2cm] &\qquad\qquad\qquad\qquad\qquad +
\bigl|\chi_{(6,m,s)}+\chi_{(12,m,s)}+\chi_{(16,m,s)}+\chi_{(22,m,s)}\bigr|^2 \,.
\eqlabel{E8}
\end{align}
Obviously, $E_6$ and $E_8$ are pure extensions, while $E_7$ is a combination
of an extension (by the same simple current as for $D_{10}$) and an exceptional
fusion-rule automorphism.

To be sure, these are not all modular invariants of the bosonic coset model 
\eqref{mimocoset}, see \cite{gannon9} for a complete list. For instance, 
one can imagine dividing out the $A_{k+1}$ model by an arbitrary subgroup of 
$\ZZ_h$. Certainly, this will give a modular-invariant partition function of
simple-current type. However, only for $\ZZ_2$ (generated by $\J$) does
the spectral-flow operator survive the projection (in other words, $q_\L\neq
q_\R\bmod\ZZ$ otherwise). Therefore, these modular invariants are usually not 
considered in the context of $\N=2$ minimal models.

Another simple modification we can do to the above modular invariants is
``orbifolding by $(-1)^F\;$''. Again, this can be understood as a simple-current
invariant, with the simple current $(0,0,2)$. For instance, the so-modified 
$A_{k+1}$ invariant reads
\begin{equation}
Z^{A_{k+1},(-1)^F} = \sum_{(l,m,s),\;s\;{\rm even}}\bigl| \chi_{(l,m,s)} \bigr|^2
+ \sum_{(l,m,s),\;s\;{\rm odd}}\chi_{(l,m,s)}\chibar_{(l,m,-s)} \,.
\eqlabel{At}
\end{equation}
Since the parity of $s$ distinguished NS and R sectors, we see that the effect
of this modification is simply to reverse the chirality of the R sector. In
particular, while the supersymmetric index ${\rm tr}(-1)^F$ in the $A_{k+1}$ 
model \eqref{A} is $k+1$, it is simply $-(k+1)$ for the modified model \eqref{At}. 
Generally, one does not bother to distinguish the two models, but we will see 
in the next subsection that twists of this sort become relevant for $(\N=2)/\ZZ_2$.

Finally, we summarize the Landau-Ginzburg superpotentials associated with each of
these models in table \ref{tab:mimoclass}. All these potentials are
quasihomogeneous, and we have ``stabilized'' them by adding suitable quadratic
pieces \cite{vawa}.
 
\begin{table}[t]
\begin{center}
\begin{tabular}{lll}
modular invariant & level $k$ &LG superpotential $W$ \\
\hline
$A_n$, $n=1,2,\ldots$ & $n-1$ & $x^{n+1} + y^2+z^2$ \\
$D_n$, $n=3,4,\dots$ & $2(n-2)$ & $x^{n-1} + xy^2 + z^2$ \\
$E_6$ & $10$ & $x^3 + y^4 + z^2$ \\
$E_7$ & $16$ & $x^3 + xy^3 + z^2$ \\
$E_8$ & $28$ & $x^3 + y^5 + z^2$ 
\end{tabular}
\caption{Classification of $\N=2$ minimal models}
\label{tab:mimoclass}
\end{center}
\end{table}

\subsection{Real Landau-Ginzburg for $(\N=2)/\ZZ_2$}
\label{rlg}

Consider an $\N=2$ Landau-Ginzburg theory, with action
\begin{equation}
\cals(\Phi,\Phibar) = \int\de{2}{z}\de{4}{\theta} \,K(\Phi,\Phibar) + 
\Bigl(\int\de{2}{z}\de{2}{\theta} W(\Phi) + {\rm c.c.} \Bigr)\,,
\eqlabel{LG}
\end{equation}
where $\Phi$ (lowest component $\phi$) is some collection of $\N=2$ chiral 
superfields. The essence of the effective LG description is that while the 
K\"ahler potential $K$ is not protected from renormalization, the 
superpotential $W$ is invariant under RG flow.

Assume now that there is an antiholomorphic involution $\omega$ that is a 
symmetry of the action, in other words,
\begin{equation}
\omega:\Phi\mapsto\omega(\Phi)=\omegat(\Phibar)\,,\;\text{$\omegat$ holomorphic,
such that}\; W(\omega(\Phi))= \Wbar(\Phibar)\,.
\end{equation}

We want the orbifold of \eqref{LG} by $\omega$. In the course of the
construction, we encounter twisted partition functions. For instance, the torus
partition function with $\omega$-twist in space direction on the worldsheet is
given by the path-integral
\begin{equation}
\gsq{\id}{\omega} = \int\limits_{\genfrac{}{}{0pt}{2}{\!\!\!\!\!\Phi(\sigma=2\pi)=}
{\quad=\omega(\Phi(\sigma=0))}}
\De\Phi\De\Phibar \ee^{\ii\cals(\Phi,\Phibar)} \,.
\eqlabel{stwist}
\end{equation}

The simplest object to calculate in such a theory is the supersymmetric Witten 
index ${\rm tr}(-1)^F$, see \cite{witten10}. In the $\omega$-twisted sector, 
this index is given by the path-integral \eqref{stwist} with periodic
boundary conditions on the fermions in both time and space direction on the
worldsheet. Actually, since the index is invariant under deformations that do
not change the singularity structure, we can calculate it in a semiclassical
approximation, see section 10 of \cite{witten10}. Explicitly, this means that
we deform $W$ by adding suitable mass terms, in such a way that the deformed
potential is still invariant under $\omega$. Then we look for fermionic zero
modes, localizing the path-integral near the critical points of the potential.

The $\omega$-twisted ground states are, in addition, localized near the fixed
points of the involution, $\Phi=\omega(\Phi)$. In linear approximation, $\omega$ 
divides the complex fields $\phi$ into a real part, $\Re(\phi)\equiv\phir$, which is 
invariant under $\omega$, and an imaginary part, which is inverted, 
$\omega:\phii\mapsto -\phii$. We then see that both $\phir$ and its superpartner,
a Majorana fermion $\psi$, have periodic boundary conditions around the spacelike 
circle. This allows for fermionic zero modes,
\begin{equation}
\begin{split}
\hat\psi^1&=\frac{1}{2\pi}
\int\nolimits_{\sigma=0}^{2\pi}{\rm d}\sigma\,\psi^1(\tau=0,\sigma) \\
\hat\psi^2&=\frac{1}{2\pi}
\int\nolimits_{\sigma=0}^{2\pi}{\rm d}\sigma\,\psi^2(\tau=0,\sigma) \,.
\end{split}
\eqlabel{fermzero}
\end{equation}
On the other hand, $\phii$ and its superpartner have antiperiodic boundary
conditions around the spacelike circle. They have no zero modes and hence a 
unique ground state. In other words, they are frozen. Since the superpotential 
respects the antiholomorphic involution, the fixed value $\phii=0$ is also a 
critical point of the superpotential. From now on we will drop $\phii$ from 
the calculation of ground states.

In a semiclassical approximation, the fields $\phir$ will minimize the bosonic
potential, \ie, they will go to a critical point of the real superpotential
\begin{equation}
\Wr = W|_{\phi=\omega(\phi)=\phir} \,.
\end{equation}
If the critical points of $\Wr$ are all non-degenerate, all classical
ground states are massive vacua. The bosonic and fermionic fluctuations
cancel and we are left with the sign of the Hessian. In other words, as 
in \cite{witten10}, we find the supersymmetric index to be equal to the
Morse index of $\Wr$,
\begin{equation}
\twin = \sum_{\del\Wr=0}\sgn{\det\del^2\Wr} \,.
\eqlabel{morsein}
\end{equation}

\subsection{A-type}

We now apply the results of the previous subsection to the ADE series of
minimal models. Let us concentrate for the moment on the A-series, with
complex superpotential $W(x) = x^h$. Obviously, the antiholomorphic
symmetries are
\begin{equation}
\omega_M: x \mapsto \ee^{2\pi\ii M/h} \, \xbar \,,
\eqlabel{omM}
\end{equation}
for $M=0,1,\ldots,h-1$. The fixed planes, to which the twisted path integral 
\eqref{stwist} localizes, are given by $x = \ee^{2\pi\ii M/2h}\, \xr$, with 
$\xr$ real. Thus, the real superpotential is
\begin{equation}
\Wr(\xr) = (-1)^M \xr^h \,.
\eqlabel{WrA}
\end{equation}

After deforming the superpotential to resolve the singularity, for instance by 
$\Wr\mapsto\Wr+\xr$, we then find for the supersymmetric index in the twisted 
sector
\begin{equation}
\twin = \sum_{\del\Wr=0}\sgn{\det\del^2\Wr} = \begin{cases} 
0 &\qquad\text{$h$ odd} \\
(-1)^M &\qquad\text{$h$ even}
\end{cases}
\eqlabel{Atwin}
\end{equation}

To see how the real LG potential captures the conformal field theory, 
recall that the Landau-Ginzburg field $x$ corresponds in the minimal 
model to the chiral-chiral primary $\phi_{(1,1,0)}^\L\phi_{(1,1,0)}^\R\,$,
while $\xbar$ corresponds to the antichiral-antichiral primary
$\phi_{(1,-1,0)}^\L\phi_{(1,-1,0)}^\R\,$. We conclude that the action
of $\omega_M$, eq.\ \eqref{omM}, in the conformal limit has to
become $\phi_{(1,1,0)}^\L\phi_{(1,1,0)}^\R\mapsto \ee^{2\pi\ii M/h}
\phi_{(1,-1,0)}^\L\phi_{(1,-1,0)}^\R\,$. The only way in which this
can be a symmetry of the conformal field theory is if the general
action is 
\begin{equation}
\phi_{(l,m,s)}^\L\phi_{(l,m,s)}^\R\mapsto \ee^{2\pi\ii mM/h} \,
\phi_{(l,-m,-s)}^\L\phi_{(l,-m,-s)}^\R\,. 
\eqlabel{ncact}
\end{equation}
We see that $\omega_M$ differs from the chiral action that we have considered 
in the previous section, eq.\ \eqref{primauto}, just by a phase factor 
$\ee^{2\pi\ii mM/h}$. This phase factor can also be expressed in terms of the 
U$(1)$ charge $q$. Namely, the phase is just $\ee^{2\pi\ii qM}$ in the NSNS
sector and $\ee^{2\pi\ii (q+1/2)M}$ in the RR sector.

Consequently, the partition function for the $\ZZ_2$ orbifold of the $A_{k+1}$ 
model by \eqref{omM} contains, besides the untwisted term \eqref{A}, the 
twisted contribution
\begin{equation}
\gsq{\omega_M}{\id} =
\sum_{} \ee^{2\pi\ii mM/h} \bigl|\chi^{(0)}_{(l,m,s)}\bigr|^2 \,.
\eqlabel{gsqM1}
\end{equation}
Here, the sum is over all symmetric fields (\ie, those fixed under
\eqref{primauto}) and the $\chi^{(0)}$ are the corresponding twining characters
(see section \ref{mimo} and appendix \ref{chzto}). Now all symmetric sectors
have $m=0$ or $h$ except $(k/2,h/2,\pm 1)$, which occurs only for $k$ even.
So the phase factor influences the partition function only if $k$ is even, and
in this case only the parity of $M$ matters, just as for the real superpotential
\eqref{WrA}.

In the untwisted sector, the effect of the phase factor is essentially
to multiply the neutral Ramond ground state (which is fixed in the
chiral $\ZZ_2$ action) by $(-1)^M$. Thus, the neutral ground state
is projected out for $M$ odd and kept for $M$ even. In other words,
the index in the untwisted sector is
\begin{equation}
\unin = \begin{cases}
\frac{k+1}2 &\quad \text{$k$ odd} \\
\frac{k}{2} + \frac12\bigl(1+(-1)^M\bigr) &\quad \text{$k$ even}
\end{cases} 
\end{equation}

The modular S-transformation of \eqref{gsqM1} yields the twisted
sectors. It turns out that the twisted sector is diagonal independently
of $M$, except for the twisted RR ground state, which occurs
in the chiral sector $\tw(k/2,0,+)\plus\,$. Namely, we find
\begin{equation}
\textstyle\frac12\displaystyle
\Bigl(\gsq{\id}{\omega_M} + \gsq{\omega_M}{\omega_M}\Bigr) = 
\sum_{(\lambda,\sigma)\psi;\;\lambda<k/2} 
\bigl|\chi_{\tw(\lambda,\sigma)\psi}\bigr|^2 + 
\sum_{\sigma,\eta,\psi} 
\chi_{\tw(k/2,\sigma,\eta)\psi} \chibar_{\tw(k/2,\sigma,(-1)^M\eta)\psi} 
\eqlabel{Atwpf}
\end{equation}
{} From \eqref{Atwpf}, one reads off that $\twin$ coincides precisely with 
the Morse index \eqref{Atwin} of the real LG potential.

Thus, we have seen that for even $k$, there are two modular invariants,
corresponding to the two possible real forms of the simple singular functions.
In the twisted sector, the difference between the two is the chirality of the
Ramond ground state. It is correlated with the orbifold action on the
neutral ground state in the untwisted sector, and can be traced back in the
LG picture to the phase in \eqref{omM}. 

For completeness, we mention that the difference between these two possibilities
can be understood as a simple-current modification of the modular invariant for 
$\mimoorb_k$. This is in fact the easiest way to check modular invariance. Using 
the fusion rules derived from the S-matrix obtained in section \ref{mimo}, one 
can check that the relevant simple-current group is $\ZZ_2\times\ZZ_2$, generated 
by $\us(0,0,2)\plus$ and $\us(0,0,2)\minus\,$. 

\subsection{D-type}
\label{Dtype}

As we have reviewed above, the D-type models can be thought of as $\ZZ_2$ 
orbifolds of the A-series. In the LG setup, we start from the $A_{k+1}$ potential 
$\tilde W=\tilde x^h$, and orbifold by $\J:\tilde x\mapsto -\tilde x$. As it 
turns out, this orbifold, which is a special case of those considered in 
\cite{vafa4}, can be effectively described by the D-type LG potential $W(x,y,z)=
x^{h/2}-xy^2-z^2$. Here, $x=\tilde x^2$ is the invariant untwisted field and $y$ 
is in the $\J$-twisted sector. We have added a quadratic stabilization term $z^2$ 
and chosen the signs in $W$ for later convenience, but of course this is irrelevant
at this stage. We now want to mod out this $D_{k/2+2}$ model by an additional 
$\ZZ_2$ that acts as an antiholomorphic involution. 

Let us first give a convenient parametrization of the antiholomorphic involutions
that fix $W$. Recall that in the A-type model, we could twist the action of
$\omega$ by the phase factor $\ee^{2\pi\ii q}$, where $q$ is the U$(1)$ charge.
In the D-type models, the LG fields $x$, $y$, and $z$ have U$(1)$ charge $2/h$,
$k/2h$, and $1/2$, respectively. This can be read off from the scaling property
\begin{equation}
W(\lambda^{2/h}x, \lambda^{k/2h}y, \lambda^{1/2}z) = \lambda W(x,y,z) \,.
\end{equation}
Consequently, we write the antiholomorphic involutions for $D_{k/2+2}$ as
\begin{equation}
\omega:\qquad
x\mapsto \ee^{2\pi\ii\, 2M/h} \xbar \qquad 
y\mapsto (-1)^{\Xi_y} \ee^{2\pi\ii\, kM/2h} \ybar \qquad
z\mapsto (-1)^{\Xi_z} \ee^{2\pi\ii\, M/2} \zbar \,,
\eqlabel{DomM}
\end{equation}
where $M\in \{0,\ldots,h-1\}$, and the remaining freedom is parametrized by the
additional signs $(-1)^{\Xi_y}$ and $(-1)^{\Xi_z}$. Note that for $k/2$ even, 
$(M,(-1)^{\Xi_y},(-1)^{\Xi_z})$ is equivalent to $(M+h/2,(-1)^{\Xi_y},
-(-1)^{\Xi_z})$, while for $k/2$ odd, $(M,(-1)^{\Xi_y},(-1)^{\Xi_z})$ is 
equivalent to $(M+h/2,-(-1)^{\Xi_y},(-1)^{\Xi_z})$.

{} From the point of view of the original $A_{k+1}$ model, the resulting
model is a $\ZZ_2\times\ZZ_2$ orbifold, and we have the possibility
of turning on discrete torsion \cite{vafa}. Since $y$ is in the $\J$-twisted
sector, and discrete torsion manifests itself in relative phases between
differently twisted sectors, we can identify this $\ZZ_2\times\ZZ_2$
discrete torsion with $(-1)^{\Xi_y}$. This will also be the interpretation
from the conformal field theory point of view.

Solving for the fixed points under \eqref{DomM}, we find that the
$\omega$-twisted sector is governed by the real superpotential
\begin{equation}
\Wr(\xr,\yr,\zr) = (-1)^M \bigl( \xr^{h/2} - (-1)^{\Xi_y} \xr\yr^2 -
(-1)^{\Xi_z} \zr^2 \bigr)\,.
\eqlabel{WrD}
\end{equation}

We are now ready to compute the index in the twisted sector. The result depends
on the phases and on $k/2$ being even or odd. We find, after a suitable
resolution of the singularity,
\begin{multline}
\twin = \sum_{\del\Wr=0}\sgn{\det\del^2\Wr} = \\
\begin{array}{c|cccc}
\bigl((-1)^{M+\Xi_z},(-1)^{\Xi_y}\bigr) & (+,+) & (+,-) & (-,+) & (-,-) \\
\hline 
\text{$\frac k2$ even} & 2 & 0 & -2 & 0 \\
\text{$\frac k2$ odd}  & 1 & 1 & -1 & -1
\end{array}
\eqlabel{Dtwin}
\end{multline}

This pattern is compatible with the classification of real singular
functions up to a real change of variables and up to stable equivalence
(\ie, adding extra variables with purely quadratic potential).

For instance, if $k/2$ is odd, we can remove $(-1)^{\Xi_y}$ in \eqref{WrD}
by redefining $\xr\mapsto -\xr$, so the two functions are equivalent over
the reals. The index is independent of $(-1)^{\Xi_y}$. If $k/2$ is even, on
the other hand, we cannot remove the relative sign between $\xr^{h/2}$
and $\xr\yr^2$ by a real change of variables, and the index depends on
$(-1)^{\Xi_y}$.

Furthermore, only the overall sign $(-1)^{M+\Xi_z}$ in front of the quadratic 
term in $\Wr$ affects the index, and we shall henceforth set $(-1)^{\Xi_z}=1$. 
Note that without the quadratic piece, the index would be independent of 
$(-1)^M$, but once a single quadratic piece is present, this classification of 
real singular functions is stable. We can add more quadratic variables to the 
potential if we so desire, but the index only depends on the overall signature 
of the quadratic part.


As before, all of these LG potentials can be related to a particular modular
invariant for the conformal field theory. To see how this works if $k/2$ is 
even, we rewrite the ($\omega$-)untwisted partition function \eqref{Deven} as
\begin{equation}
\gsq{\id}{\id} = \sum_{(l,m,s),\;l\;{\rm even}} \Bigl(
\bigl|\chi_{(l,m,s)}\bigr|^2 + \chi_{(l,m,s)} \chibar_{(k-l,m,s)} \Bigr)\,.
\end{equation}
With $\omega$-twist in time direction on the worldsheet, we have
\begin{equation}
\gsq{\omega}{\id} = \sum_{(l,m,s),\;l\;{\rm even}} \ee^{2\pi\ii mM/h}\Bigl( 
\bigl|\chi^{(0)}_{(l,m,s)}\bigr|^2 + (-)^{\Xi_y}
\chi^{(0)}_{(l,m,s)} \chibar^{(0)}_{(k-l,m,s)}
\Bigr)\,,
\eqlabel{gsqDo}
\end{equation}
where we have immediately inserted the phase choices corresponding to
those in the involution \eqref{DomM}. For $\ee^{2\pi\ii mM/h}$, this
can be justified through the U$(1)$-charge just as for the A-type
models. The sign $(-1)^{\Xi_y}$ is the relative phase between
untwisted and $\J$-twisted sectors, and hence is clearly discrete torsion.

The signs appear in particular in the action of $\omega$ on the untwisted
neutral Ramond ground states. Namely, there are two such ground states in 
\eqref{Deven}, on which $\omega$ is represented as
\begin{equation}
\begin{pmatrix}
(-1)^{M} & 0 \\ 0 & (-1)^{M+\Xi_y}
\end{pmatrix} \,.
\end{equation}
So, the index in the untwisted sector of the orbifold is
\begin{equation}
\unin = \frac k4 + \frac12\,\bigl(2 + (-1)^M + (-1)^{M+\Xi_y}\bigr)\,.
\end{equation}

After modular transformation, we then find the contribution from 
the $\omega$-twisted sectors,
\newcommand\topa[2]{\genfrac{}{}{0pt}{2}{\scriptstyle #1}{\scriptstyle #2}}
\begin{equation}
\begin{split}
&\hskip 3.1in\bigl((-1)^M,(-1)^{\Xi_y}\bigr) \\
{\textstyle \frac12}\Bigl(\gsq{\id}{\omega} + \gsq{\omega}{\omega}\Bigr) &= 
\left\{
\begin{array}{lc}
\displaystyle\sum_{\topa{(\lambda,\sigma)\psi}{\lambda\;{\rm even}}}
\hskip -0.2cm
2\,\bigl|\chi_{\tw(\lambda,\sigma)\psi}\bigr|^2 \,+\, 
\displaystyle\sum_{\sigma,\eta,\psi}
2\,\bigl|\chi_{\tw(k/2,\sigma,\eta)\psi}\bigr|^2
& (+,+) \\[0.8cm]
\displaystyle\sum_{\topa{(\lambda,\sigma)\psi}{\lambda\;{\rm odd}}}
\hskip -0.2cm
2\bigl|\chi_{\tw(\lambda,\sigma)\psi}\bigr|^2  & (+,-) \\[0.8cm]
\displaystyle\sum_{\topa{(\lambda,\sigma)\psi}{\lambda\;{\rm even}}}
\hskip -0.2cm
2\,\bigl|\chi_{\tw(\lambda,\sigma)\psi}\bigr|^2 \!+
\!
\!\!\sum_{\sigma,\eta,\psi}\!\!
2\chi_{\tw(k/2,\sigma,\eta)\psi} \chibar_{\tw(k/2,\sigma,-\eta)\psi} 
& (-,+) \\[0.8cm]
\displaystyle\sum_{\topa{(\lambda,\sigma)\psi}{\lambda\;{\rm odd}}} 
\hskip -0.2cm
2\,\bigl|\chi_{\tw(\lambda,\sigma)\psi}\bigr|^2 & (-,-)
\end{array}
\right.
\end{split}
\eqlabel{Detwpf}
\end{equation}
Again, the index in the twisted sector coincides with the LG result 
\eqref{Dtwin}. 

{} From the point of view of $\mimoorb_k$, these modular invariants can
be understood as follows. The model with $\twin=2$ is a simple
simple-current extension by the simple current $\us(k,0,0)\plus\,$. If
we extend by $\us(k,0,0)\minus\,$, we project onto $\lambda$ odd
in the twisted sector, and obtain the model with $\twin=0$. Finally,
the model with $\twin=-2$ is obtained from the first one by using the 
simple-current group $\ZZ_2\times\ZZ_2$ generated by $\us(0,0,2)\plus$ 
and $\us(0,0,2)\minus\,$, similarly to the A-type models.

For $k/2$ odd, we can similarly construct two different D-type models
based on $\mimoorb_k$. Namely, we can use either of the simple currents
$\us(k,0,0)\psi$, with $\psi=\pm$, to form a modular invariant (the twist 
by the $\ZZ_2\times\ZZ_2$ leads to nothing new). We obtain the twisted 
sector contribution
\begin{equation}
\begin{split}
&\hskip 3.725in\; \psi \\
{\textstyle \frac12}\Bigl(\gsq{\id}{\omega} + \gsq{\omega}{\omega}\Bigr) &=
\left\{\hspace{-.1in}
\begin{array}{lc}
\displaystyle\sum_{(\lambda,\sigma)\psi,\;\lambda<k/2}
\hspace{-.2in}
\bigl|\chi_{\tw(\lambda,\sigma)\psi}\bigr|^2  \!+\!
\sum_{\sigma,\eta,\psi} \,
\bigl|\chi_{\tw(k/2,\sigma,\eta)\psi}\bigr|^2 &  + \\[0.6cm]
\displaystyle\sum_{(\lambda,\sigma)\psi,\;\lambda<k/2}
\hspace{-.2in}
\bigl|\chi_{\tw(\lambda,\sigma)\psi}\bigr|^2 \, \!+\!  
\sum_{\sigma,\eta,\psi}
\chi_{\tw(k/2,\sigma,\eta)\psi} \chibar_{\tw(k/2,\sigma,-\eta)\psi} &  -
\end{array}
\right.
\end{split}
\eqlabel{Dotwpf}
\end{equation}
Comparing this with the two possibilities in \eqref{Dtwin}, we see that
we have to identify $\psi=(-1)^M$. The part of the partition function that
corresponds to $\gsq{\omega}{\id}$ is given by
\begin{equation}
\gsq{\omega}{\id} = \sum_{(l,m,s),\;l\;{\rm even}}  
\bigl|\chi^{(0)}_{(l,m,s)}\bigr|^2 + \psi \;
\chi^{(0)}_{(k/2,h/2,s)} \chibar^{(0)}_{(k/2,h/2,s)} \,,
\end{equation}
and the index in the untwisted sector is
\begin{equation}
\unin = \frac{k+2}{4} + \frac12(1+ \psi) \,.
\end{equation}
We note that from \eqref{DomM}, we would have naively expected $\omega$ to
act with $(-1)^{M+\Xi_y}$, and not with $\psi=(-1)^M$, on the neutral RR 
ground state, which is in the $\J$-twisted sector. However, we can notice
that the action with $(-1)^M$ is the only one that is compatible with the 
identifications given below eq.\ \eqref{DomM}.

\subsection{E-type}

The possible real forms of the exceptional E-type LG potentials are
as follows
\begin{equation}
\begin{array}{lll|ll}
\hbox to 0.4in {} & W       & \Wr            & \unin & \twin   \\\hline
E_6 & x^3+y^4+z^2 \quad\qquad& \xr^3\pm\yr^4\mp\zr^2 \quad\qquad & 3& 0 \\
E_7 & x^3+xy^3+z^2& \xr^3+\xr\yr^3\mp\zr^2 & 3+\frac12(1\pm1)\quad&\pm1\\
E_8 & x^3+y^5+z^2 & \xr^3+\yr^5\mp\zr^2    & 4     & 0
\end{array}
\eqlabel{Etwin}
\end{equation}
We note that for $E_6$, there are two real forms that are not equivalent
to each other by a real change of variables, yet they are not distinguished 
by any index. It would be interesting to see whether there is any other 
signature in the LG theory that distinguishes the two. Furthermore, we
note that the only model which does have twisted sector ground states is
$E_7$. This is also the only one that has a neutral ground state in the
untwisted sector.

It is natural to expect that there is a modular invariant of $(\N=2)/\ZZ_2$
associated with each of the models in \eqref{Etwin}. Since the explicit
forms are quite complicated for notational reasons, we shall describe our
findings in words.

Recall that the $E_6$ modular invariant of the $\N=2$ minimal model at level
$10$ is an exceptional extension by the field $(6,0,0)$. In the orbifold,
and in analogy with the $D_{\rm even}$ invariants, one would then expect to 
be able to extend $\mimoorb_{10}$ by either of $\us(6,0,0)\plusminus\,$. 
However, it turns out that this is not the case. Only the extension by 
$\us(6,0,0)\minus$ is possible. This is presumably related to the fact 
that $(6,0,0)$ is not a simple current, and as a consequence the extension 
of $\omega$ to the extended chiral algebra of the $E_6$ model is more
restricted.

Something similar happens for the $E_7$ invariant. We could well extend 
$\mimoorb_{16}$ by either of $\us(16,0,0)\plusminus\,$, and this leads to the
possibilities for $D_{10}$. But only the extension by $\us(16,0,0)\minus$ has 
an exceptional fusion rule automorphism that could correspond to the $\ZZ_2$ 
orbifold of the $E_7$ modular invariant. We have checked explicitly all modular 
invariants related to $E_6$ and $E_7$. While we have not done so for $E_8$, we 
do not expect any real surprises there.

\section{$G_2$-orbifolds in the Landau-Ginzburg phase}
\label{full}

We now return to the $G_2$-holonomy geometries $X=\frac{Y\times S^1}
{\ZZ_2}$. Let us summarize our results so far. We have described some general 
aspects of these geometries in section \ref{geoinv}. In section \ref{lomo}, we 
have discussed the local geometry around the singularities, which is a singular
$A_1$ fibration over a supersymmetric three-cycle $M$. It turned out that 
topological twisting does not fully specify this fibration, and that additional 
data in the form of a real line bundle $L$ over $M$ is needed. We have then seen 
in section \ref{glomo} how this line bundle is determined by the global B-field
configuration on the Calabi-Yau space, and we have computed the low-energy spectrum
of strings in some relevant examples of hypersurfaces, 
\begin{equation}
Y = \bigl\{\textstyle \sum x_i^{h_i}=0 \bigr\} \subset \PP_{w_1,\ldots,w_5}^4 \,.
\end{equation}

As is well-known, in the small-volume region of moduli space, the sigma model on 
$Y$ is described by a Landau-Ginzburg orbifold, with superpotential
\begin{equation}\eqlabel{lgsuperpota}
W=\sum_{i=1}^5 x_i^{h_i}\,,
\end{equation}
and $\ZZ_h$ orbifoldization generated by
\begin{equation}
u:x_i\mapsto \ee^\frac{2\pi \ii w_i}{h}x_i \,,
\eqlabel{uLG}
\end{equation}
where $h=\lcm(h_i)$ and $w_i=\frac{h}{h_i}$.

The additional $\ZZ_2$ (antiholomorphic involution) that yields $G_2$-holonomy
acts on the LG fields by
\begin{equation}
\omega:x_i\mapsto \rho_i\bar x_i \,,
\eqlabel{omegaLG}
\end{equation}
as discussed in section \ref{geoinvglsm} using the GLSM.

In section \ref{lgorbi}, it was argued that the massless spectrum in the twisted
sector of such an orbifold is given by the Morse index of the appropriate real 
section of the LG potential. This was then illustrated for the ADE minimal models. 
We now apply these ideas to the LG phase of our $G_2$-geometries.

\subsection{A nonabelian Landau-Ginzburg orbifold}

The actions of $\ZZ_h$, \eqref{uLG}, and $\ZZ_2$, \eqref{omegaLG}, do not commute 
with each other. The generators satisfy the relation
\begin{equation}
u\omega=\omega u^{-1} \,.
\end{equation}
So, the group generated by $u$ and $\omega$ is not $\ZZ_h\times\ZZ_2$, but rather
the dihedral group, $D_h$. This group can be visualized as the group of rotations
and reflections of an $h$-gon in the plane\footnote{The dihedral group should not
be confused with the binary dihedral group, which is one of the ADE subgroups of
SU$(2)$.}.

The full Landau-Ginzburg model\footnote{The $S^1$ component of $X$ does not
significantly affect the present discussion, since it is frozen in the twisted
sectors we are about to discuss.} is therefore a non-abelian orbifold of
\eqref{lgsuperpota}. According to the usual prescription \cite{dhvw,ginsparg},
such an orbifold receives contributions from twists by any commuting pair $g_1$, 
$g_2$ of elements of the orbifold group, \ie, the torus amplitude is
\begin{align}
Z_{\rm orb}&=\frac{1}{|D_h|}\sum_{\genfrac{}{}{0pt}{2}{g_1,g_2\in D_h}{[g_1,g_2]=0}}
\gsq{g_2}{g_1} \,, \\
\intertext{%
where $|D_h|=2h$ is the order of the dihedral group. In other words, there is one
twisted sector for each conjugacy class $C_{g_1}$ of $D_h$, and the projection in
each sector is by summation over the stabilizer group, $N_{g_1}$.}
Z_{\rm orb}&=\sum_{C_{g_1}\subset D_h}\frac{1}{|N_{g_1}|}
\sum_{g_2\in N_{g_1}}\gsq{g_2}{g_1} 
\end{align}
For $D_h$, $h$ even, the conjugacy classes and the corresponding stabilizer groups 
are listed in table \ref{concla}. For $h$ odd, the table looks a little different,
but all partition functions we will compute vanish anyway in that case.

\begin{table}
\begin{center}
\begin{tabular}{|c|c|c|}
\hline
$C_{g_1}$            & $g_1$                 &  $N_{g_1}$ \\
\hline
$\{\id\}$             & $\id$                 &  $D_h$ \\
$\{u^\frac{h}{2}\}$  & $u^\frac{h}{2}$       &  $D_h$ \\
$\{u^m,u^{-m}\}$     & $u^m$                 &  $\ZZ_h=\{u^n\}$ \\
$\{\omega,\omega u^{2},\ldots\}$     & $\omega$  &
$\{\id,u^\frac{h}{2},\omega,\omega u^\frac{h}{2}\}$ \\
$\{\omega u,\omega u^3,\ldots\}$ & $\omega u$ &
$\{\id,u^\frac{h}{2},\omega u,\omega u^{\frac{h}{2}+1}\}$ \\
\hline
\end{tabular}
\end{center}
\caption{Conjugacy classes of $D_h$ and their representatives and stabilizer 
groups.} 
\label{concla}
\end{table}

In section \ref{glomo}, we have obtained the spectrum of massless strings in
the $\omega$-twisted sector. In the LG description, this includes the classes 
$C_\omega$ and $C_{\omega u}$. The corresponding partition functions read
\begin{equation}
\begin{split}
Z_\omega &= \frac{1}{4}\Bigl(\gsq{\id}{\omega}+\gsq{\omega}{\omega}+
\gsq{u^\frac{h}{2}}{\omega}+\gsq{\omega u^\frac{h}{2}}{\omega}\Bigr) \\
&=\frac{1}{2}\Bigl(\gsq{\id}{\omega}+\gsq{u^\frac{h}{2}}{\omega}\Bigr) \,,
\end{split}
\eqlabel{pfo} 
\end{equation}
where in the second line we used that ground states are invariant under
a modular T-transformation, and
\begin{equation}
Z_{\omega u}=
\frac{1}{2}\Bigl(\gsq{\id}{\omega u}+\gsq{u^\frac{h}{2}}{\omega u}\Bigr) \,,
\eqlabel{pfou}
\end{equation}
respectively. These partition functions are be evaluated with periodic boundary
conditions on the worldsheet supercurrents in spacelike direction, in order to
be in the RR sector, and also with periodic boundary conditions on the worldsheet 
supercurrents in timelike direction in order to get a topologically protected 
quantity, the Witten index. But exactly what do these quantities count?

\subsection{Indices and duality}
\label{indices}

In general, the Witten index does not quite give the number of ground
states in a supersymmetric field theory. Rather, it only counts the
ground states weighted with $(-1)^F$. In the context of a sigma model
with target space $M$, say, ground states correspond to cohomology classes
\cite{witten10}, and the Witten index
\begin{equation}
{\tr}(-1)^F = \sum (-1)^i\, b_i = \chi(M)
\end{equation}
is equal to the Euler characteristic of the target space. While it is
expected \cite{witten11} that also the total number of ground states is
equal to the dimensionality of the total cohomology, this can not be
deduced from $\tr(-1)^F$ alone. For example, on a compact $G_2$-manifold, 
like on any other odd-dimensional manifold, the index is zero, even though
we certainly do not expect supersymmetry to be spontaneously broken.
Fortunately sometimes, there are finer indices that can be used to put
constraints on the number of ground states being lifted up in pairs.

For instance \cite{witten10}, for a sigma model into an $N$-dimensional
sphere, $\tr(-1)^F=1+(-1)^N$ vanishes in odd dimensions. But the sphere has
an isometry $L:S^N\rightarrow S^N$ that inverts one of the $N+1$ coordinates
of $\RR^{N+1}\supset S^N$, and this gives rise to a symmetry of the sigma
model. Implementing $L$ on the cohomology, one finds that the corresponding
index, called Lefshetz index, is $\tr L(-1)^F= 1-(-1)^N$. So the total number 
of ground states is always $2$---as long as $L$ is preserved, of course.

In the context of Calabi-Yau compactifications, it is the U$(1)$ R-symmetry
on the worldsheet, with current $J=\ii\sqrt\frac{c}{3}\partial\phi$, that
allows to understand many properties of the nonlinear sigma model \cite{lvw}.
In particular, the left- and right-moving U$(1)$ charges of Ramond ground
states are directly equal to the holomorphic and antiholomorphic degrees of
the corresponding cohomology elements.

For sigma models into $G_2$-manifolds, there is neither a conserved U$(1)$
current nor in the generic case an isometry that one could use to tighten the
links between the cohomology and the space of ground states. But $G_2$ sigma
models are conformal field theories with extended chiral algebras. And luckily,
it turns out that there is a $\ZZ_2$ symmetry, very much analogous to $L$ for 
the $N$-sphere, whose index actually gives the total number of ground states. 

To define $L$, let us recall the extended chiral algebra associated with $G_2$ 
holonomy \cite{shva,figueroa}. This chiral algebra is generated by three bosonic 
and three fermionic fields, $(T,X,K)$ and $(G,\Phi,M)$. Here, $(T,G)$ generate
an $\N=1$ superconformal algebra with central charge $\frac{21}2$, $X$ and $\Phi$ 
generate an $\N=1$ superconformal algebra with central charge $\frac{7}{10}$ 
(the tri-critical Ising model), and $K$ and $M$ are the superpartners (with
respect to $G$) of $\Phi$ and $X$, respectively. We refer to the above references
for the details of this algebra.

As in any $\N=1$ supersymmetric theory, the $G_2$-holonomy algebra has the
symmetry
\begin{align}
(-1)^F& : \quad (T,X,K,G,\Phi,M)\mapsto (T,X,K,-G,-\Phi,-M)
\eqlabel{minus}\\
\intertext{%
that gives a sign to all fermionic generators. Furthermore, one can read off
from the formulae in \cite{shva,figueroa} that there is an additional $\ZZ_2$
symmetry that acts as}
L&:\quad (T,X,K,G,\Phi,M) \mapsto (T,X,-K,G,-\Phi,M) \,.
\eqlabel{L}
\end{align}
Thus, $L$ commutes with the supersymmetry charge, and we can define the
``Lefshetz index'', $\tr L (-1)^F$. 

To explain what this index counts, we need some more information on the ground 
states. Recall from \cite{shva} that the zero mode $\Phi_0$ of $\Phi$ acts almost
like Poincar\'e duality. More precisely, from the algebra it follows that on the 
ground states, $\Phi_0$ and its right-moving counterpart $\Phibar_0$ form a 
two-dimensional Clifford algebra. Since $(-1)^F$ anti-commutes with both $\Phi_0$ 
and $\Phibar_0$, it is identified with the chirality operator. The smallest 
irreducible representation of this algebra is two-dimensional, and the two ground 
states have opposite eigenvalue of $(-1)^F$. They can be considered dual to each 
other\footnote{On a manifold of $G_2$ holonomy, the associative three-form acts
almost like the Poincar\'e duality operator on the cohomology already at the level 
of classical geometry. This follows by decomposing the cohomology according to
$G_2$ representations and using Clifford multiplication of forms. One can also 
derive certain other aspects of the tri-critical Ising model from these 
considerations. We will discuss this elsewhere.}. So $\Phi_0$ and $\Phibar_0$ 
pair up ground states with opposite $(-1)^F$, and $\tr(-1)^F$ vanishes.

These considerations are completely identical to those for the supersymmetric
system of a free boson and fermion on a circle, see, for instance, \cite{ginsparg}.
There, the fermion zero modes $\psi_0$ and $\bar\psi_0$ form a two-dimensional 
Clifford algebra which is represented on two ground states with opposite $(-1)^F$.
Note that in both cases, $(-1)^F$ is the non-chiral fermion number.

Now by definition \eqref{L}, $L$ squares to $1$, leaves $T$ and $G$ invariant, and
anti-commutes with $\Phi_0$ and $\Phibar_0$ (in a non-chiral representation).
Therefore, it has to take eigenvalues $\pm 1$ on the two ground states that are
paired by $\Phi_0$. In other words, $L$ is equal to $(-1)^F$ on the ground states.
Of course it is not equal to $(-1)^F$ in general, since otherwise we could not
use it to define our index. As a consequence, $\tr L(-1)^F$ counts the
total number of ground states.
\footnote{%
We would like to mention at this point that the index $\tr L(-1)^F$ could be used 
to give a new and very much simplified proof of the fact that marginal operators 
in $G_2$-holonomy CFT are exactly marginal \cite{shva}. Namely, by spectral flow, 
there is a one-to-one correspondence between RR ground states and marginal 
operators in the NSNS sector. Since by the index $\tr L(-1)^F$, ground states 
cannot be lifted by deformations that preserve $G_2$ holonomy, all marginal 
operators must remain marginal. This will be discussed more fully elsewhere.}

The analog of $L$ for the supersymmetric $S^1$ is the isometry we have
mentioned above in the example of the $N$-dimensional sphere. It inverts the
fermion, but, unlike $(-1)^F$, also inverts the boson, and so commutes with the
supercharge.

While it is remarkable that such an operator exists for any $G_2$-holonomy CFT,
for our $G_2$-manifolds which are orbifolds of Calabi-Yau spaces times a circle,
there is even more. Let us denote by $I$ the quantum $\ZZ_2$ symmetry (the simple
current) that is dual to $\omega$. This symmetry is broken as soon as we move
away from the orbifold point, but at the orbifold we can define indices like
$\tr I(-1)^F$ and even $\tr L I(-1)^F$. These indices can be used to disentangle
the ground states from different sectors in the orbifold.

We are now finally in a position to explain what we are counting in the LG
orbifold. Recall from the results of \cite{figueroa} the origin of the operator 
$\Phi$ in the Calabi-Yau model. From geometry, we know the expression \eqref{assform} 
for the associative three-form, and this is the natural, and as it turns out the
correct, ansatz for the CFT operator. Now the Calabi-Yau part of $\Phi$ contains
the spectral flow operator, and in the LG orbifold, spectral flow is implemented
by an additional $u$-twist. We conclude that the ground states in \eqref{pfo} and
\eqref{pfou} are related to each other by acting with the zero mode $\Phi_0$. In 
other words, they are Poincar\'e dual to each other and in particular,
$Z_\omega+Z_{\omega u}=0$. Moreover, we note that if all ground states in 
$Z_\omega$ have the same chirality, this can be expressed by $|Z_\omega| = \tr\, 
\frac14 (1-I) L (-1)^F$.

There is a conveninent way to visualize these different twisted sectors.
If we consider both, the left- and the right-moving sector, the U$(1)$ current
algebra in the Calabi-Yau part can be thought of as a free boson $\phi$ on a 
circle of radius $\sqrt{c/3}$, as recently exploited in \cite{douglascat}. 
Since the antiholomorphic involution acts on $J$ as $J\mapsto -J$, we get an 
orbifold theory which ``contains'' a free boson on an interval. Then there is the
additional $S^1$ factor in $X=\frac{Y\times S^1}{\ZZ_2}$. The involution has
$4$ fixed points on the two circles, and $4$ twisted sectors. As a consequence,
we have a fourfold degeneracy of ground states. One factor of two corresponds
to the doubling of the fixed point set $M$ of $\omega$ in $Y\times S^1$,
and the remaining degeneracy to Poincar\'e duality.

\subsection{An ambiguity?}

We now turn to the computation of the indices, \eqref{pfo} and \eqref{pfou}, 
in the LG orbifold. The basic idea is that, as in explained in section
\ref{lgorbi}, the twist by $\omega$ in \eqref{omegaLG} together with the
periodic boundary conditions on the worldsheet supersymmetry currents freezes the 
imaginary parts of the $x_i$'s and we are left with the real LG potential,
\begin{equation}
\Wr=\sum_{i=1}^5 \eta_i\xr_i^{h_i}\,,
\eqlabel{fullrealLG}
\end{equation}
for the real parts $\xr_i=\Re \bigl(\rho_i^{-1/2}x_i\bigr)$. Here, the signs 
$\eta_i=\rho_i^{h_i/2}$ are the same as in \eqref{eqrealhys}. We could then 
analyse this potential in the framework of supersymmetric quantum mechanics to 
determine the ground states.

Equivalently, we can see the ground states in a semi-classical approximation. We
deform the superpotential \eqref{lgsuperpota} by mass terms in a way that respects 
the involution and makes the critical point of the superpotential nondegenerate,
\begin{equation}
W=\sum_{i=1}^5 (x_i^{h_i}+m_i\rho_i^{-1}x_i^2) \,,
\end{equation}
where the $m_i$'s are real.

As before, we have a separation into real and imaginary parts of $x_i$. The
imaginary parts have no zero modes, and are frozen, while the real parts have 
fermionic zero modes $\hat\psi_i^1$ and $\hat\psi_i^2$, as in \eqref{fermzero}.
The real superpotential then becomes
\begin{equation}
\Wr=\sum_{i=1}^5\bigl(\eta_i\xr_i^{h_i}+m_i\xr_i^2\bigr) \,.
\end{equation}

\begin{figure}
\begin{center}
\psfrag{pota}{$\CW=\xi^h$}
\psfrag{potb}{$\CW=\xi^h+\xi^2$}
\psfrag{potc}{$\CW=\xi^h-\xi^2$}
\epsfig{file=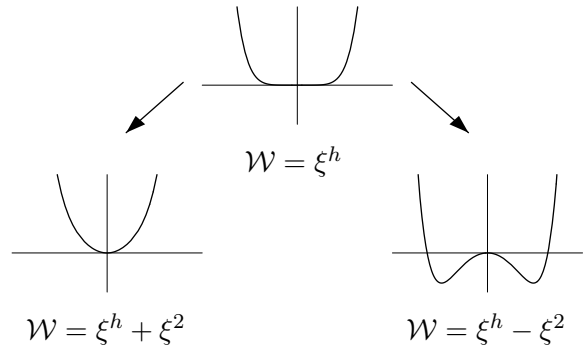,width=8cm}
\caption{The two deformations of the real superpotential.}
\label{morse}
\end{center}
\end{figure}

If we now choose the signs of $m_i$ to be $\eta_i$, we get exactly one critical 
point of $\Wr$ for $\xr_i=0$ (see figure \ref{morse}) and we can neglect the 
fluctuations of the boson. The fermionic zero modes satisfy the anti-commutation 
relations of a Clifford algebra and the Hamiltonian for the fermion zero modes 
can be written as
\begin{equation}
H=\sum_i\bigl(-\ii m_i\hat\psi_i^1\hat\psi_i^2\bigr) \,.
\end{equation}

This Hamiltonian has a unique ground state in the $2^5$-dimensional representation
of the Clifford algebra. Namely, it picks out the state for which each
$\ii\hat\psi_i^1\hat\psi_i^2$ has the eigenvalue $\eta_i$. Since the worldsheet
fermion number operator has the form
\begin{equation}
(-)^F=\prod_i\bigl(\ii\hat\psi_i^1\hat\psi_i^2\bigr) \,,
\eqlabel{minusF}
\end{equation}
we obtain the first bit of \eqref{pfo},
\begin{equation}
\gsq{\id}{\omega}=\prod_i\eta_i \,.
\end{equation}
Note that this is the Morse index for the real superpotential $\Wr$.

To determine $\gsq{u^{\frac h2}}{\omega}$, all we need to do is to represent the
operator $u^\frac{h}{2}$ in our Clifford algebra. Now it follows from \eqref{uLG}
that $u^{\frac h2}$ acts on the fields $\xr_i$ and $\psi_i$ by $(-1)^{w_i}$. Hence a 
representation of $u^{\frac h2}$ in the Clifford algebra is given by
\begin{equation}
u^\frac{h}{2}= \pm\prodprime_i{}\bigl(\ii\hat\psi_i^1\hat\psi_i^2\bigr) \,,
\eqlabel{uh2cliff}
\end{equation}
where $\prodprime$ indicates a product over those $i$ with $w_i$ odd. The sign
ambiguity arises because if $u^{\frac h2}$ satisfies $u^{\frac h2}\psi_i 
u^{\frac h2}=-\psi_i\,$, the negative of $u^{\frac h2}$ will satisfy the same 
equation. In other words, we only know how $u^{\frac h2}$ acts on fields, but its 
action on states is ambiguous. However, this is an ambiguity that arises only 
once, say for the ground state of the Clifford algebra representation. Once this 
is fixed, the action on the remaining states is determined, since the $\psi_i$'s 
generate the representation, and we know how $u^{\frac h2}$ acts on those. Below, 
we will fix this overall sign ambiguity by a single comparison with geometry. 

Inserting \eqref{uh2cliff} into the trace gives
\begin{equation}
\gsq{u^\frac{h}{2}}{\omega}=\pm\Bigl(\prod_i\eta_i\Bigr)
\Bigl(\prodprime_i\eta_i\Bigr) \,.
\end{equation}
Thus, we obtain
\begin{equation}\eqlabel{lgtwspectrum}
Z_\omega=\frac 12\prod_i\eta_i \Bigl( 1\pm\prodprime_i{}\eta_i\Bigr) \,.
\end{equation}

Turning now to the $(\omega u)$-twisted sector, one can see from the definitions
\eqref{uLG} and \eqref{omegaLG} that the $(\omega u)$-twisted sector is governed
by the real superpotential $(-\Wr)$, \ie, the negative of \eqref{fullrealLG}.
All the other operators are unchanged\footnote{To be precise, this is naively
the case only for an odd number of LG variables, which is the natural description 
of a Calabi-Yau threefold. If there is an even number of variables, for instance 
after addition of more quadratic terms to the potential, $u$ has to be defined 
as anticommuting with $(-1)^F$. In other words, we would have to change the 
definition of $(-1)^F$, eq.\ \eqref{minusF}, in this sector.}. As a consequence, 
$Z_{\omega u}= -Z_\omega$, as expected from the general considerations in section 
\ref{indices}.

Finally, let us mention that we can also construct the explicit modular-invariant
partition functions corresponding to the $\ZZ_2$ orbifolds of Gepner models
times a free boson and fermion. Namely, using the results of sections \ref{mimo}
and \ref{lgorbi}, we can compute the non-abelian orbifold of a tensor product
of $\N=2$ minimal models. We leave the explicit form of these partition functions
for a future work, and just point out the relevant features. Whenever at least one 
level in the Gepner model is even, the $\ZZ_h$ Gepner orbifold includes the 
extension by a simple current $\J\cong u^{\frac h2}$ generating a $\ZZ_2\subset\ZZ_h$. 
The orbifold, then, can be extended either by $\J\plus$ or by $\J\minus\,$. As in 
the D-type modular invariants, this implies different projections in the twisted 
sector (namely, the constraint $\sumprime \lambda_i$ even or odd, respectively).
In particular, when all levels are even, the tensor product has massless 
Ramond ground states, which are kept or projected out in the $G_2$-holonomy model, 
just as from Landau-Ginzburg.

\subsection{Comparison with geometry}

We now fix the ambiguity in the ``parity of the ground state'' that appears in
the definition of $u^{\frac h2}$, eq.\ \eqref{uh2cliff}, by ``comparison with 
experiment''. Observe that if all $\eta_i$ are $+1$, the geometric fixed point 
set $M$ in the large-volume limit is empty, and there are no massless twisted 
strings. We conclude that we have to choose the minus sign. While the naturality
of this choice leads us to believe that the sign is uniquely fixed, we do not 
know at this stage whether this can be done in a less {\it ad hoc} way.

However, once we have fixed this sign ambiguity once and for all, we can compare
the massless spectra computed in the LG phase with those from section \ref{glomo}.
And indeed, we find perfect agreement. When the geometry predicts absence of
massless modes, the LG result is zero. (This is also so for the quintic. Although
we assumed throughout this section that all variables appear with even powers,
it is easy to see in the LG setup that when there is at least one odd power,
the index vanishes.)

To understand that the spectra also match when there actually are massless modes
in the twisted sector, recall from the local or the global model that $\hat b_0(M)$
counts the number of (twisted-sector) vector multiplets in three dimensions, and 
$\hat b_1(M)$ the number of chiral multiplets. In other words, $\hat b_0(M)$ acts 
as an ``effective $b_2$'' of the $G_2$-manifold, while $\hat b_1(M)$ acts as an
``effective $b_3$''. If, as it happens, a non-vanishing $\hat b_0$ is replaced
with a non-vanishing $\hat b_1$, we expect the chirality of the corresponding 
ground state to be flipped. In the LG phase, this is reflected by the fact that
$Z_\omega$ and $Z_{\omega u}$ change sign when in the large-volume phase
$\hat b_0=1$ is replaces with $\hat b_1=1$. Using $\tr L(-1)^F$, we can write
\begin{equation}
\tr \;\frac14 (1-I) L (-1)^F = \hat b_0 + \hat b_1 = \pm Z_\omega = 
\mp Z_{\omega u} \,.
\eqlabel{pfooo}
\end{equation}

Again, there is a sign ambiguity in \eqref{pfooo}, because we cannot decide
from the LG whether $\hat b_0$ or $\hat b_1$ is non-zero. However, this sign 
ambiguity is physical. It is simply a reflection of the fact that if stringy 
effects are taken into account, we cannot really distinguish between $b_2$ and 
$b_3$ of a $G_2$-manifold. In other words, this is a reflection of mirror 
symmetry for $G_2$-manifolds \cite{shva}. Two different classical geometries 
give rise, in the stringy regime, to one and the same conformal field theory. 
One can also test directly that the conformal field theories are equivalent 
from explicit Gepner-model calculations. And indeed, using methods similar to 
those in section \ref{lgorbi}, we have found just enough Gepner-model partition 
functions to match the massless spectra in the twisted sector that we have found 
here, but not as many as one would have expected from the possibilities for the 
geometric involutions. So we conclude that these geometries must be mirror to 
each other.

Once all ambiguities are removed, the formula \eqref{lgtwspectrum} reproduces 
exactly the spectrum given in tables \ref{mm11114} and \ref{mm11222} in section
\ref{glomo}. We also found agreement for other simple cases of Calabi-Yau 
hypersurfaces with all $h_i$ even. However, the determination of the fixed 
point set of the anti-holomorphic involutions becomes increasingly complex for 
models that have more than one K\"ahler modulus, see appendix \ref{sslag} for an 
example.

\begin{acknowledgments}
We have greatly benefited from discussions with
Bobby Acharya,
Per Berglund,
Andi Brandhuber,
Alex Buchel,
Iouri Chepelev,
Rich Corrado,
Oliver DeWolfe,
Mike Douglas,
Jaume Gomis,
Shamit Kachru,
Calin Lazaroiu,
Wolfgang Lerche,
Joe Polchinski,
Martin Ro\v{c}ek,
Nick Warner,
and Ed Witten,
and we would like to thank them for their valuable suggestions.
The research of R.R.\ was supported in part by the DOE under Grant No.\
91ER40618(3N) and by the National Science Foundation under Grant No.\
PHY00-98395(6T). The research of C.R.\ was supported in part by the DOE
under the Grant No.\ DE-FG03-84ER-40168. The research of J.W.\ was supported 
in part by the National Science Foundation under Grant No.\ PHY99-07949. 
R.R.\ and J.W.\ would like to thank the CIT-USC Center for Theoretical 
Physics for hospitality during the work on this project.
\end{acknowledgments}

\begin{appendix}

\section{Special Lagrangians in $\PP_{11222}^4[8]$}
\label{sslag}

In this appendix we analyze special Lagrangian submanifolds of the 
Calabi-Yau hypersurface $\PP_{11222}^4[8]$ which arise in the construction
of $G_2$-manifolds as fixed point sets of anti-holomorphic involutions. 
This example is somewhat complicated because the embedding space has a 
complex-codimension 2 singularity and thus the generic Calabi-Yau hypersurface 
will be singular. Techniques for dealing with this type of singularities are 
described, for example, in \cite{kmp}. In the language of the GLSM one blows 
up the singularity by introducing an additional chiral multiplet and an 
additional U$(1)$ gauge field that can be used to gauge the former to a 
constant. This amounts to replacing the singular points of the CY space 
with $\PP^2$'s with size equal to the FI term of the new gauge field.

As in the examples described in section \ref{glomo}, different choices
of anti-holomorphic involution lead to different fixed point sets. We
will begin with a description of facts that are independent of the signs
in the real section of the CY surface and then present the details for the
choice that leads to a geometry and topology of the fixed point set that is
(very) different from the ones discussed in the main text.

Special Lagrangians in the GLSM context have also been discussed in 
\cite{leva,gjs15,hiv,agva,gjs3}. These constructions are related to, but not
identical with, ours.

\subsection{Construction}

To determine the geometry of the real section of $\PP_{11222}^4[8]$, we need
to intersect the real homogenous hypersurface equation
\begin{equation} \tag{1}
\xi_6^4(\eta_1\xi_1^8+\eta_2\xi_2^8)+
\eta_3\xi_3^4+\eta_4\xi_4^4+\eta_5\xi_5^4=0
\end{equation}
in $\RR^6$ with the two real D-flatness conditions
\begin{align}
\xi_1^2+\xi_2^2-2\xi_6^2 &=r^2>0 \,, \tag{2} \\
\xi_3^2+\xi_4^2+\xi_5^2+\xi_6^2 &=R^2>0 \,, \tag{3}
\end{align}
and then divide out by the residual $\ZZ_2\times\ZZ_2$ gauge 
symmetry which is the part of the original U$(1)\times {\rm U}(1)$ 
gauge symmetry that preserves reality properties of various fields. 
The two D-flatness conditions pick exactly one $\ZZ_2\times\ZZ_2$ orbit 
in $\RR^6$ for each point in the 'real toric variety'. Since we are only
interested in the topology of the cycle, any other two hypersurfaces in 
$\RR^6$ with the same property would be equally good.

Since equation $(1)$ contains fourth powers of the $\xi_i$, more 
convenient (real) D-flatness conditions are
\begin{align}
\xi_1^4+\xi_2^4-2\xi_6^4 &=r^4>0 \,, \tag{$2'$} \\
\xi_3^4+\xi_4^4+\xi_5^4+\xi_6^4 &=R^4>0 \,.\tag{$3'$}
\end{align}
Let us procede by first solving $(2')$ for $\xi_6$. For a solution 
to exist, $\xi_1$ and $\xi_2$ have to satisfy the inequality
\begin{equation}
\xi_1^4+\xi_2^4\ge r^4 \,, \tag{$*$}
\end{equation}
which constrains $(\xi_1,\xi_2)$ to lie outside a deformed circle of 
'radius' $r^4$. If $(*)$ is satisfied, we can solve $(2')$ for $\xi_6$; 
the solution is double valued in the allowed range of $\xi_1$ and $\xi_2$, 
except at the inner boundary (circle) where the two branches meet and 
$\xi_6=0$. The picture in the $(\xi_1,\xi_2,\xi_6)$-space is a throat 
geometry. Using the solution for $\xi_6$ in $(3')$ we find
\begin{equation}
\xi_1^4+\xi_2^4+2\xi_3^4+2\xi_4^4+2\xi_5^4=2R^4+r^4>0 \,. \tag{$3''$}
\end{equation}

Before introducing the new U$(1)$ gauge field and its FI parameter the 
point $\xi_1=\xi_2=0$ was a singular point containing a circle shrinking 
to zero size. We see that the blow-up of $\PP_{11222}^4[8]$ removed this 
singular point and created the throat geometry.

Since each $\eta_i$ takes only two values, $\pm 1$, there will always be
two of the last three terms in equation $(1)$ that will have the same sign. 
Let us then assume, without loss of generality, that $\eta_4=\eta_5$. Then 
all equations only contain $\xi_4^4+\xi_5^4$. We therefore see that the 
solution to all three equations $(1)$, $(2)$, and $(3)$, is a topologically 
trivial circle fibration over some two dimensional surface $\Sigma$ in the 
four-dimensional $(\xi_1,\xi_2,\xi_3,\xi_6)$-space. The fiber might shrink 
to zero size at some boundaries of $\Sigma$.

Equation $(3'')$ has solutions for the radius of the fiber, $\xi_4^4+
\xi_5^4$, if 
\begin{equation}
\xi_1^4+\xi_2^4+2\xi_3^4\le 2R^4+r^4 \,. \tag{$**$}
\end{equation}
This constrains the solutions to a finite region of the
$(\xi_1,\xi_2)$-plane. When the equality is saturated the fiber 
shrinks to zero size.

\begin{figure}[ht]
\begin{center}
\psfrag{x4}{$\scriptstyle\xi_3=0$}
\psfrag{x6}{$\scriptstyle\xi_6=0$}
\psfrag{x3}{$\scriptstyle\xi_4^4+\xi_5^4=0$}
\psfrag{xo}{$\scriptstyle\xi_1^4+\xi_2^4=2R^4+r^4$}
\epsfig{file=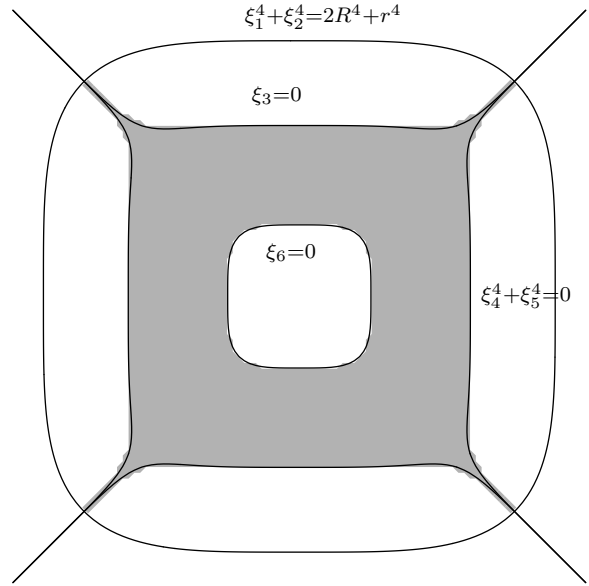,width=8cm}
\caption{The region $A$ in the $(\xi_1,\xi_2)$-plane, for 
$(\eta_1,\eta_2,\eta_3,\eta_4,\eta_5)=(1,-1,-1,1,1)$.\label{Aregion}}
\end{center}
\end{figure}

We can now insert the solutions of $(2')$ and $(3'')$ into the hypersurface
equation $(1)$ and solve for $\xi_3$. This will necessarily introduce another
double cover of the $(\xi_1,\xi_2)$-plane coresponding to positive and negative 
$\xi_3$. These two branches meet at points where $\xi_3$ vanishes. The precise 
solution depends of course on the $\eta_i$. The inequality resulting from
\begin{equation}
\xi_3^4 \ge 0 \,, \tag{$*{*}*$}
\end{equation}
together with $(*)$ and $(**)$ then specify a certain region $A$ in the 
$(\xi_1,\xi_2)$-plane (figure \ref{Aregion}). At boundaries where $(*)$ or 
$(*{*}*)$ are saturated, there are throats to other branches of the 4-fold 
covering of $A$. Finally, at boundaries where $(**)$ is saturated, the circle 
fiber shrinks to zero size.

\begin{figure}
\begin{center}
\psfrag{a}{$a$}
\psfrag{b}{$b$}
\psfrag{c}{$d$}
\psfrag{d}{$c$}
\psfrag{e}{$e$}
\psfrag{f}{$f$}
\epsfig{file=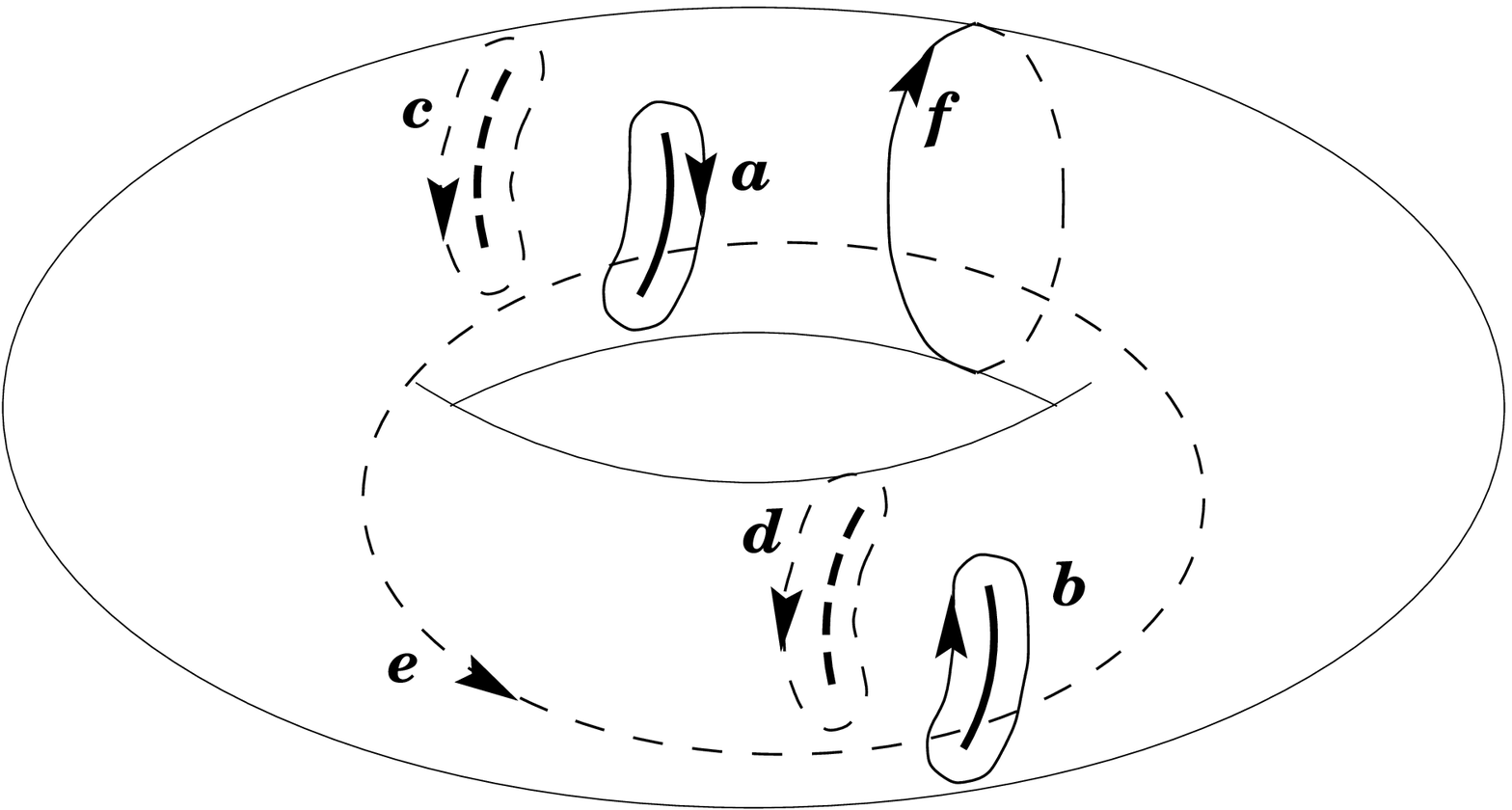,width=8cm}
\caption{$\Sigma$ is a 2-torus with four holes. The holes are 
shown as cuts. We also display representatives of homology 
1-cycles.}
\label{torus}
\end{center}
\end{figure}

To determine the geometry more accurately, let us specialize to
$(\eta_1,\eta_2,\eta_3,$ $\eta_4,\eta_5)=(1,-1,-1,1,1)$.
This seems to be the most interesting case which we were not able 
to solve by simpler methods. Inserting equations $(2')$ and $(3'')$
in equation $(1)$ we find
\begin{equation}
\textstyle \half(\xi_1^4+\xi_2^4-r^4)(\xi_1^8-\xi_2^8)-
\half(\xi_1^4+\xi_2^4)-2\xi_3^4+R^4+\half r^2=0 \,, \notag
\end{equation}
and the constraints become
\begin{align*}
(*)   \quad&\Rightarrow\qquad& \xi_1^4+\xi_2^4 & \ge r^4 \,, \\
(**)  \quad&\Rightarrow\qquad& (\xi_1^4+\xi_2^4-r^4)(\xi_1^8-\xi_2^8)+
(\xi_1^4+\xi_2^4) &\le 2R^4+r^4 \,, \\
(*{*}*) \quad&\Rightarrow\qquad& (\xi_1^4+\xi_2^4-r^4)(\xi_1^8-\xi_2^8)-
(\xi_1^4+\xi_2^4) &\ge-2R^4-r^4 \,.
\end{align*}
The solution to those constraints is shown in figure \ref{Aregion}. 
Patching the 4-fold cover together shows that $\Sigma$ is a 2-torus with 
four holes as depicted in figure \ref{torus}. The circle is fibered over 
$\Sigma$ in the way described above.

The special lagrangian $M$ of interest now is the quotient of the
circle fibration over $\Sigma$ by the residual $\ZZ_2\times\ZZ_2$
gauge symmetry. We want to understand the topology of this 3-manifold.

\subsection{Topology}

We can construct representatives of the homology cycles of $M$ by
starting from the covering space, $\tilde M$, which is the trivial circle
fibration over $\Sigma$. Since the fibers shrink to zero size at the
boundaries of $\Sigma$ (denoted in figure \ref{torus} by the thick lines),
there are no 1-cycles that reside in the fiber direction and therefore,
using the K\"uneth formula, the 1-dimensional homology is determined by the
1-dimensional homology of the base. One possible choice of representatives
of homology classes with orientations is shown in figure \ref{torus}. Note, 
however, that the six cycles in figure \ref{torus} form an overcomplete 
basis of $H_1(S^1\rightarrow\Sigma)$ since they satisfy the relation
\begin{equation}
a+b+c+d=0 \,.
\eqlabel{homconstraint}
\end{equation}

Finding representatives of the homology classes of $M$ is simplified by that
the $\ZZ_2\times\ZZ_2$ action has no fixed points. Therefore, all we have to
do is to trace the original homology basis through the $\ZZ_2\times\ZZ_2$
projection and keep the invariant cycles.

By the charge table \eqref{charget} in section \ref{withblowup}, the first $\ZZ_2$ 
acts by $(-1,-1,1,1,1,1)$ on the six coordinates. This just halves the size of the 
$f$ cycle and identifies the cycles $a$ and $b$ with cycles $d$ and $c$, 
respectively. The second $\ZZ_2$ acts with charges $(1,1,-1,-1,-1,-1)$. 
Therefore, it will halve the size of the $e$ cycle, identify cycles $a$ and 
$d$ with $c$ and $b$, respectively, as well as shrink each circle fiber to half 
of its original size. Table \ref{ztzt} shows the action of the two $\ZZ_2$'s 
on the chosen (overcomplete) basis.

\begin{table}[ht]
\begin{center}
\begin{tabular}{|c|c|c|}
\hline
 & $(-1,-1,1,1,1,1)$ & $(1,1,-1,-1,-1,-1)$ \\
\hline
$a$ & $d$ & $c$ \\
$b$ & $c$ & $d$ \\
$c$ & $b$ & $a$ \\
$d$ & $a$ & $b$ \\
$e$ & $e+b+d$ & $e$ \\
$f$ & $f$ & $f+a+d$ \\
\hline
\end{tabular}
\caption{$\ZZ_2\times\ZZ_2$ action on the cohomology of $\tilde M$.}
\label{ztzt}
\end{center}
\end{table}

Since the two $\ZZ_2$ actions commute, they can be simultaneously
diagonalized. Using the constraint equation \eqref{homconstraint}, one shows
that there are two homology classes that are invariant under the action of
$\ZZ_2\times\ZZ_2$. Thus, $b_0(M)=1$ and $b_1(M)=2$. Furthermore, there
is exactly one class in each non-trivial representation of $\ZZ_2\times
\ZZ_2$. These results are summarized in table \ref{ztztt}.

\begin{table}[ht]
\begin{center}
\begin{tabular}{|c|c|c|}
\hline
 & $(-1,-1,1,1,1,1)=+1$ & $(-1,-1,1,1,1,1)=-1$\\
\hline
$(1,1,-1,-1,-1,-1)=+1$ & 
\hbox{\lower7pt\hbox{$
\stackrel{\displaystyle{2f+a+d}}{\displaystyle{2e+b+d}}$}} & $a-b+c-d$ \\
\hline
$(1,1,-1,-1,-1,-1)=-1$ & $a+d$ & $a+b-c-d$ \\
\hline
\end{tabular}
\caption{Eigenvectors of the $\ZZ_2\times\ZZ_2$ action on the cohomology.}
\label{ztztt}
\end{center}
\end{table}

With these results in hand, we can now also compute the twisted
(co)homology discussed in section \ref{glomo}. Recall that the real line 
bundle $L$ over $M$ is given by $L = \frac{\tilde M\times\RR}{\ZZ_2\times
\ZZ_2}$, where the $\ZZ_2\times\ZZ_2$ acts on the $\RR$ fiber in a certain
representation determined by the global B-field. The twisted cohomology 
is simply the ordinary cohomology of $\tilde M$ that is in the same 
representation as the fiber. Table \ref{tb11222} gives these Betti
numbers for all possible twists.

\begin{table}[t]
\begin{center}
\begin{tabular}{|cc|c|cc|cc|cc|cc|}
\cline{4-11}
\multicolumn{3}{c|}{} &
\multicolumn{8}{c|}{$\ZZ_2\times\ZZ_2$ representation of fiber}   \\
\multicolumn{3}{c|}{} &
\multicolumn{2}{c|}{$(+,+)$} &
\multicolumn{2}{c|}{$(+,-)$} &
\multicolumn{2}{c|}{$(-,+)$} &
\multicolumn{2}{c|}{$(-,-)$} \\\hline
\#($\eta_1,\eta_2=-$) & $\!$\#($\eta_3,\eta_4,\eta_5=-$) & $\tilde M$ &
$\hat b_0$ & $\hat b_1$ &
$\hat b_0$ & $\hat b_1$ &
$\hat b_0$ & $\hat b_1$ &
$\hat b_0$ & $\hat b_1$ \\\hline
0 & 0 & $\emptyset$                & 0 & 0 & 0 & 0 & 0 & 0 & 0 & 0 \\
0 & 1 & $S^1\times S^2\times S^0$  & 1 & 1 & 1 & 1 & 0 & 0 & 0 & 0 \\
0 & 2 & $S^1\times S^1\times S^1$  & 1 & 3 & 0 & 0 & 0 & 0 & 0 & 0 \\
0 & 3 & $S^1\times S^2\times S^0$  & 1 & 1 & 1 & 1 & 0 & 0 & 0 & 0 \\
1 & 0 & $S^3\times S^0\times S^0$  & 1 & 0 & 1 & 0 & 1 & 0 & 1 & 0 \\
1 & 1 & $S^1$ over $\Sigma$        & 1 & 2 & 0 & 1 & 0 & 1 & 0 & 1 \\
\hline
\end{tabular}
\caption{The twisted Betti numbers of the fixed cycle in $\PP_{11222}^4[8]$,
depending on the $\eta_i$ and the $\ZZ_2\times\ZZ_2$ twist. The first
$\ZZ_2$ acts only on $\xi_1$ and $\xi_2$, the second on $\xi_3$, $\xi_4$,
$\xi_5$, and $\xi_6$.}
\label{tb11222}
\end{center}
\end{table}

The results of the LG phase are consistent with the spectra in the
last two columns of this table, which is what appears in table
\ref{mm11222}.

\section{Chiral $\ZZ_2$ orbifolds}
\label{chzto}

In this appendix, we summarize from refs.\ \cite{bifs,dvvv} some 
generalities about $\ZZ_2$ orbifolds of rational conformal field theories 
originating from an order $2$ automorphism of the chiral algebra.
\nxt
Let $\cala$ be a rational chiral algebra with irreducible representations
$(R_\lambda,\calh_\lambda)$ labelled by $\lambda\in\Lambda$. Let
$\omega$ be an order $2$ automorphism of $\cala$, and $\calao$ the
subalgebra of $\cala$ that is left pointwise fixed under $\omega$.
\nxt
By definition, for every representation $R_\lambda$, $R_\lambda\chain\omega$
is again a representation of $\cala$. In this way, $\omega$ induces an action
on the set of irreducible representations, $\omega^*:\Lambda\to\Lambda$.
Assume that $\omega$ can be implemented as a twisted intertwiner on the
representation spaces, \ie,
\begin{equation}
\calt^\omega: \calh_\lambda\to\calh_{\omega^*\lambda} \text{\; such that for 
all $a\in\cala$,\; } \calt^\omega\chain R_\lambda(\omega(a)) = R_{\omega^*\lambda}(a)
\chain\calt^\omega \,.
\eqlabel{autoh}
\end{equation}
\nxt
Any irreducible representation $\lambda$ of $\cala$ induces by restriction a
representation of $\calao$. This representation is irreducible precisely if
$\omega^*\lambda\neq\lambda$ (non-symmetric representation). The representations 
induced from $(R_\lambda,\calh_\lambda)$ and $(R_{\omega^*\lambda},
\calh_{\omega^*\lambda})$ then become isomorphic and give rise to one irreducible 
representation of $\calao$. If, on the other hand, $\omega^*\lambda= \lambda$ 
(symmetric representation), then the induced representation is reducible (and 
fully reducible) as a representation of $\calao$, and gives rise to two
irreducible representations of $\calao$. Those irreducible representations
which are induced from representations of $\cala$ are said to be ``in the
untwisted sector''.
\nxt
There are furthermore irreducible representations which are not induced from
those of $\cala$. They are said to be ``in the twisted sector''. 
\nxt
To summarize, there are three different kinds of irreducible representations of 
$\calao$. Untwisted representations inherit their labels from $\cala$. If they come
from non-symmetric representations, we shall denote them by $\un\lambda\equiv\un(
\omega^*\lambda)$. The two irreducible representations that come from symmetric 
representations of $\cala$ will be distinghuished by labels $\us\lambda\plus$ and 
$\us\lambda\minus$. Finally, twisted representations require new labels, of the 
generic form $\tw\dot\lambda\psi$, with $\psi=\pm$.
\nxt
According to \cite{dvvv,bifs}, the full modular data of the orbifold can be
determined from the so-called twining characters. Formally, these are 
defined for every symmetric representation $\lambda$ by the formula
\begin{equation}
\chi^{(0)}_{\lambda}(2\tau)={\rm tr}_{\calh_\lambda} \calt^\omega q^{L_0-c/24}\,,
\end{equation}
where, as usual, $q=\ee^{2\pi\ii\tau}$, and $\calt^\omega:\calh_\lambda
\to\calh_\lambda$ is as in eq.\ \eqref{autoh}.
\nxt
The characters of the orbifold theory in the untwisted sector are then given by
\begin{align}
\chi^\calo_{\un\lambda}(\tau) &= \chi_{\lambda}(\tau) 
\eqlabel{chiun} \\
\chi^\calo_{\us\lambda\psi}(\tau) &= \frac{1}{2}\bigl( \chi_{\lambda}(\tau)
+\psi\,\eta_\lambda^{-1}\, \chi^{(0)}_{\lambda}(2\tau) \bigr) \,.
\eqlabel{chius}
\end{align}
Here, $\eta_\lambda$ is a certain conventional phase, which is determined
by the conjugation properties of the representation $\lambda$. 
\nxt
Under modular transformation, the characters in \eqref{chiun} and \eqref{chius}
do not close onto themselves, but rather lead to the characters in the twisted
sectors. Namely, for every twisted sector $\dot\lambda$, there is a function 
$\chi^{(1)}_{\dot\lambda}(\tau)$ such that the characters of the two twisted
representations are given by
\begin{equation}
\chi^\calo_{\tw\dot\lambda\psi}(\tau) = \frac12\Bigl( \chi^{(1)}_{\dot
\lambda}(\frac\tau2)+\psi\, \bigl(T_{\dot\lambda}^{(1)}\bigr)^{-1/2}\,
\chi^{(1)}_{\dot\lambda}\bigl(\frac{\tau+1}2\bigr) \Bigr) \,.
\eqlabel{chitw}
\end{equation}
Here, $T_{\dot\lambda}^{(1)}$ is part of (the square of) the modular T-matrix
and determines the conformal weights in the twisted sectors up to (half) 
integers.
\nxt
The relevant modular-S transformations are
\begin{align}
\chi^{(0)}_{\lambda}\bigl(-\frac1\tau\bigr) &=
\sum_{\dot\mu} S^{(0)}_{\lambda,\dot\mu} \chi^{(1)}_{\dot\mu}(\tau) 
\eqlabel{szero} \\
\intertext{and}
\chi^{(1)}_{\dot\lambda}\bigl(-\frac1\tau\bigr) &=
\sum_\mu S_{\dot\lambda,\mu}^{(1)} \chi^{(0)}_{\mu}(\tau) \,,
\eqlabel{sone} 
\end{align}
where the matrices $S^{(0)}$ and $S^{(1)}$ are square, non-singular,
and related to each other by phases.
\nxt
We now collect from \cite{bifs} the formulae for the modular S-matrix, 
$S^\calo$, of the orbifold. Matrix elements of $S^\calo$ between only 
untwisted representations depend only on the S-matrix of the parent theory,
\begin{align}
S^\calo_{\un\lambda,\un\mu} &= S_{\lambda,\mu}+S_{\lambda,\omega^*\mu}\notag\\
S^\calo_{\un\lambda,\us\mu\psi} &= S_{\lambda,\mu} \eqlabel{soun}\\
S^\calo_{\us\lambda\psi,\us\mu\psi'} &= \frac12 S_{\lambda,\mu} \,. \notag\\
\intertext{Matrix elements between untwisted and twisted representations 
involve, of course, $S^{(0)}$ and $S^{(1)}$,}
S^\calo_{\un\lambda,\tw\dot\mu\psi} &= 0 \notag\\
S^\calo_{\us\lambda\psi,\tw\dot\mu\psi} &= \frac12\psi\eta_\lambda^{-1}
S^{(0)}_{\lambda,\dot\mu} = \frac12\psi\eta_\lambda S^{(1)}_{\dot\mu,\lambda}
\eqlabel{soustw}\\
\intertext{Finally, matrix elements between twisted representations are given
by}
S_{\tw\dot\lambda\psi,\tw\dot\mu\psi'} &= \frac12\psi\psi'
P_{\dot\lambda,\dot\mu} \,,
\eqlabel{sotwtw}
\end{align}
where $P$ is the matrix
\begin{equation}
P = \bigl(T^{(1)}\bigr)^{1/2}S^{(1)}\bigl(T^{(0)}\bigr)^2
S^{(0)} \bigl(T^{(1)}\bigr)^{1/2} \,.
\eqlabel{po}
\end{equation}

\section{$\ZZ_2$ orbifolds of SU$(2)_k$ and U$(1)_{2N}$}
\label{sutuoo}

We here record for reference the modular data of the $\ZZ_2$ orbifolds of
SU$(2)$ WZW models and of the compactified free boson.

\subsection{SU$(2)$ WZW}

The modular data of the orbifold of the SU$(2)$ WZW model at level $k$ by charge
conjugation can easily be extracted from \cite{bifs}. Note that since charge 
conjugation is an inner automorphism for SU$(2)$, all primaries are symmetric
(\ie, self-conjugate). As in the main text, we distinguish untwisted (symmetric) 
and twisted sectors by adding a prefix $\us$ and $\tw$, respectively. The list 
of primary fields of the orbifold then is as follows.

\smallskip
\begin{tabular}{llll}
\multicolumn{1}{c}{sector} & \multicolumn{2}{l}{labels and range} & conformal
weight \\
\hline
\multicolumn{1}{c}{untwisted} & \\
\cline{1-1} \\[-0.3cm]
\parbox{3cm}{} & $\us l\psi$ & $l=0,\ldots,k$ & $\frac{l(l+2)}{4h}$ \;\;
(but $\Delta_{\us 0\plus}= 1$) \\ 
\multicolumn{1}{c}{twisted} & \\
\cline{1-1} \\[-0.3cm]
& $\tw\lambda\psi$ & $\lambda=0,\ldots,k$ & $\frac{c}{24} + \frac{(k- 2\lambda)^2}{16h}
+ \frac14(1-\epsilon_\lambda\psi)$
\end{tabular}
\smallskip

\noindent
Here, $c=3k/h$ is the central charge and $h=k+2$. In the twisted sector, the
conformal weights of the two primaries with given $\lambda$ differ by $\frac12$, 
but the choice $\epsilon_\lambda=\pm 1$ is arbitrary. The choices of \cite{bifs} 
amount to $\epsilon_\lambda = (-1)^\lambda$. This is also the assignement used in 
the main text.

The modular S-matrix is given by the formulae
\begin{equation}
\eqlabel{ssuto}
\begin{split}
S_{\us l\psi,\us l'\psi'} &= \frac{1}{\sqrt{2(k+2)}} \, \sin\pi\frac{(l+1)(l'+1)}{k+2} \\
S_{\us l\psi,\tw \lambda\psi'} &= \frac{\psi\,\ii^{-l}}{\sqrt{2(k+2)}} \,
\sin\pi\frac{(l+1)(\lambda+1)}{k+2} \\
S_{\tw \lambda\psi,\tw \lambda'\psi'} &= \frac{\psi\psi'\,\ii^{-\lambda-\lambda'} \,
\ee^{-2\pi\ii k/8}}{\sqrt{2(k+2)}} \, \sin\pi\frac{(\lambda+1)(\lambda'+1)}{k+2} \,.
\end{split}
\end{equation}

\subsection{Compactified free boson}

Consider the free boson CFT and its $\ZZ_2$ orbifold by the charge conjugation automorphism
(see, for instance, \cite{ginsparg}). If the boson is compactified on a circle with rational
radius squared, the (extended) chiral algebra of the model becomes rational. In this case,
the finite number, $2N$, of primary fields of the non-orbifolded theory are labelled by the U$(1)$ 
charge, $k=0,\ldots,2N-1\bmod 2N$. Charge conjugation acts on the U$(1)$ current as $J\mapsto -J$. 
The action on primary fields is $k\mapsto -k$, so that there are two symmetric sectors, $k=0$ 
and $k=N$. Accordingly, there are two twisted sectors, with two primary fields each. Altogether,
the orbifold has the following $N+7$ primary fields.

\smallskip
\begin{tabular}{lllll}
\multicolumn{2}{c}{sector} & \multicolumn{2}{l}{labels and range} & conformal weight \\
\hline
\multicolumn{2}{c}{untwisted} & \\
\cline{1-2} \\[-0.3cm]
\parbox{3cm}{} & non-symmetric & $\un k$ & $k=1,\ldots N-1$ & $k^2/4N$ \\
& symmetric & $\us k\psi$ & $k=0,N$; $\psi=\pm$ & 
\parbox[t]{4cm}{$\Delta_{0\plus}= 0$; $\Delta_{0\minus}=1$ \\
$\Delta_{N\plus}=\Delta_{N\minus} = N/4$} \\[0.2cm]

\multicolumn{2}{c}{twisted} & \\
\cline{1-2} \\[-0.3cm]
&& $\tw\sigma\psi$ & $\sigma=0,1$; $\psi=\pm$ &
$\frac{1}{16} + \frac14(1-\psi (-1)^{N\sigma})$ 
\end{tabular}
\smallskip

\noindent
(The conventional notation \cite{ginsparg} in the twisted sector is $\sigma_{1,2}$ for
the twist fields with $\Delta = \frac1{16}$ and $\tau_{1,2}$ for those with
$\Delta=\frac{9}{16}$.)

The corresponding characters and their modular transformation properties can be found
in \cite{dvvv}. However, as is often the case in similar situations, some of the
characters coincide, and hence this does not fully determine the modular S-matrix.
There are two more constraints that can be used to fix the resulting ambiguity:
the relation $(ST)^3 = S^2$ between modular S- and T-matrices and integrality of fusion
rules. It turns out that the S-matrices given in \cite{dvvv} do not quite satisfy the
relation $(ST)^3=S^2$ in the modular group. More precisely, the S-matrix entries
involving twisted sectors actually depend on $N\bmod 4$, and not ${}\bmod 2$, as shown
in \cite{dvvv}. But the fusion rules computed in \cite{dvvv} are integral, and indeed
coincide with the ones computed from \eqref{u1os}. The full S-matrix is given by
\begin{equation}
\begin{split}
S_{\un k,\un k'} &= \sqrt{\frac{2}{N}} \,\cos \pi\frac{kk'}{N}\\
S_{\us k\psi,\un k'} &= \frac{1}{\sqrt{2N}} \,\ee^{-\pi\ii k k'/N}\\
S_{\us k\psi,\us k'\psi'} &=\frac{1}{\sqrt{8N}} \,\ee^{-\pi\ii k k'/N}\\
S_{\un k,\tw\sigma\psi} &= 0 \\
S_{\us k\psi,\tw\sigma\psi'} &= \frac{\psi}{\sqrt{8}} 
\begin{pmatrix}
1 & 1 \\ \ii^{-N} & -\ii^{-N}
\end{pmatrix} \\
S_{\tw\sigma\psi,\tw\sigma'\psi'} &= 
\frac{\psi\psi'}4
\begin{pmatrix}
\;\;\;\;1\,+\ii^{-N}     & (-1)^N - \ii^N \\
(-1)^N - \ii^N & \;\;\;\;1\,+\ii^{-N} 
\end{pmatrix}\,,
\end{split}
\eqlabel{u1os}
\end{equation}
where rows and columns in the last two lines are indexed by $k=0,N$ and $\sigma,
\sigma'=0,1$, respectively. 

\end{appendix}

\bibliographystyle{utcaps}
\bibliography{pal}

\end{document}